\documentclass[10pt,journal,compsoc]{IEEEtran}
\pdfoutput=1
\usepackage{amsmath,amsfonts, amssymb}
\usepackage{algorithm,algorithmicx}
\usepackage{array}
\usepackage[caption=false]{subfig}
\usepackage[justification=centering]{caption}
\usepackage{cite}
\usepackage{textcomp}
\usepackage{stfloats}
\usepackage{url}
\usepackage[urlcolor=black,hidelinks]{hyperref}
\usepackage{verbatim}
\usepackage{graphicx}
\usepackage{graphics,float,epsf}
\usepackage{booktabs}
\usepackage{epstopdf,calc,import}
\usepackage{tikz}
\usepackage{pgfplots}
\usepackage{enumitem}
\pgfplotsset{compat=newest}
\pgfplotsset{plot coordinates/math parser=false}
\newlength\figureheight
\newlength\figurewidth

\usepackage[noend]{algpseudocode}
\DeclareMathOperator*{\argmax}{argmax}
\usepackage{gensymb}

\newcommand\Tstrut{\rule{0pt}{2.6ex}}         
\definecolor{los}{rgb}{0.07,0.62,1.00}

\usepackage{pifont}
\newcommand{\cmark}{\color{green} \ding{51}}%
\newcommand{\xmark}{\color{red} \ding{55}}%

\usepackage{acronym}
	\newacro{mm-wave}[mm-wave]{millimeter-wave}
	\newacro{RSS}{received signal strength}
	\newacro{RSRP}{reference signal received power}
	\newacro{UE}{user}
	\newacro{UEs}{users}
	\newacro{BS}{base station}
	\newacro{BSs}{base stations}
	\newacro{gNB}{next generation node B}
	\newacro{TX}{transmitter}
	\newacro{RX}{receiver}
	\newacro{5G-NR}{5G new radio}
	\newacro{FR1}{frequency range 1}
	\newacro{FR2}{frequency range 2}
	\newacro{CDF}{cumulative distribution function}
	\newacro{CCDF}{complementary cumulative distribution function}
	\newacro{PDF}{probability density function}	
	\newacro{RLF}{radio link failure}
	\newacro{LOS}{line-of-sight}
	\newacro{NLOS}{non-LOS}
	\newacro{CIR}{channel impulse response}
	\newacro{HPBW}{half-power beamwidth}
	\newacro{EIRP}{equivalent isotropic radiated power}
	\newacro{SNR}{signal-to-noise-ratio}
	\newacro{SINR}{signal-to-interference-plus-noise-ratio}
	\newacro{INR}{interference-to-noise-ratio}
	\newacro{OFDM}{orthogonal frequency-division multiplex}
	\newacro{TDD}{time-division duplex}
	\newacro{CS}{compressive sensing}
	\newacro{AoA}{angle of arrival}
	\newacro{AoD}{angle of departure}
	\newacro{RF}{radio frequency}
	\newacro{MIMO}{multiple input multiple output}
	\newacro{SU-MIMO}{single-user multiple input multiple output}
	\newacro{MU-MIMO}{multi-user multiple input multiple output}
	\newacro{ABF}{analog beamforming}
	\newacro{CBF}{codebook-based analog beamforming}
	\newacro{HBF}{hybrid beamforming}
	\newacro{DBF}{digital beamforming}
	\newacro{ZF}{zero forcing}
	\newacro{MMSE}{minimum mean square error}
	\newacro{CSI}{channel state information}
	\newacro{SVD}{singular value decomposition}
	\newacro{URA}{uniform rectangular array}
	\newacro{SSB}{synchronization signal block}
	\newacro{CSI-RS}{channel state information-reference signal}
	\newacro{BPL}{beam pair link}
	\newacro{SDMA}{spatial division multiple access}
	\newacro{TDMA}{time division multiple access}

\begin{document}
%
%
\title{HBF MU-MIMO with Interference-Aware Beam Pair Link Allocation for Beyond-5G mm-Wave Networks}
%

\author{Aleksandar~Ichkov\IEEEauthorrefmark{1},~\IEEEmembership{}%
	Alexander~Wietfeld\IEEEauthorrefmark{2},~\IEEEmembership{}%
	Marina~Petrova\IEEEauthorrefmark{3}~\IEEEmembership{}%
        and~Ljiljana~Simi\'c\IEEEauthorrefmark{1}~\IEEEmembership{}
       
\thanks{\IEEEauthorblockA{\IEEEauthorrefmark{1}Aleksandar Ichkov and Ljiljana~Simi\'c are with the Institute for Networked Systems at RWTH Aachen University, Aachen 52072, Germany \\ (e-mail: aic@inets.rwth-aachen.de, lsi@inets.rwth-aachen.de).}}
\thanks{\IEEEauthorblockA{\IEEEauthorrefmark{2}Alexander Wietfeld was with the Institute for Networked Systems at RWTH Aachen University, Aachen, Germany. Now he is with the Chair of Communication Networks at Technical University of Munich, Munich 80333, Germany (e-mail: alexander.wietfeld@tum.de).}}
\thanks{\IEEEauthorblockA{\IEEEauthorrefmark{3}Marina Petrova is with the Institute for Networked Systems and the Mobile Communications and Computing Teaching and Research Area at RWTH Aachen University, Aachen 52072, Germany \\ (e-mail: petrova@mcc.rwth-aachen.de).}}

}




\maketitle

\begin{abstract}
	Hybrid beamforming (HBF) multi-user multiple-input multiple-output \mbox{(MU-MIMO)} is a key technology for unlocking the directional millimeter-wave (mm-wave) nature for spatial multiplexing beyond current codebook-based \mbox{5G-NR} networks. In order to suppress \mbox{co-scheduled} users' interference, HBF MU-MIMO is predicated on having sufficient radio frequency chains and accurate channel state information (CSI), which can otherwise lead to performance losses due to imperfect interference cancellation. 
	In this work, we propose \textit{IABA}, a \mbox{5G-NR} standard-compliant beam pair link (BPL) allocation scheme for mitigating spatial interference in practical HBF \mbox{MU-MIMO} networks. \textit{IABA} solves the network sum throughput optimization via either a \mbox{\textit{distributed}} or a \textit{centralized} BPL allocation using dedicated CSI reference signals for candidate BPL monitoring. 
	We present a comprehensive study of practical \mbox{multi-cell} mm-wave networks and demonstrate that HBF MU-MIMO without interference-aware BPL allocation experiences strong residual interference which limits the achievable network performance.
	Our results show that \textit{IABA} offers significant performance gains over the default interference-agnostic \mbox{5G-NR} BPL allocation, and even allows HBF MU-MIMO to outperform the fully digital MU-MIMO baseline, by facilitating allocation of secondary BPLs other than the strongest BPL found during initial access. We further demonstrate the scalability of \textit{IABA} with increased gNB antennas and densification for beyond-5G mm-wave networks.
\end{abstract}


\bstctlcite{IEEEexample:BSTcontrol}

%

\section{Introduction}

	The spectrum-rich millimeter-wave (mm-wave) bands are a key enabler for the capacity enhancement of 5G and Beyond-5G cellular networks~\cite{xiao_millimeter_2017}, with 3GPP designating numerous bands in the \ac{FR2} between 24.5--71~GHz for commercial deployments~\cite{etsi_nr}. Overcoming the challenging propagation and blockage-prone nature of the mm-wave channel necessitates large antennas and beamforming at the \ac{gNB} and the \ac{UEs}, and a new type of beam management, in which the \ac{gNB} and UE need to sweep their beam candidates sequentially in order to establish connectivity via a directional \ac{BPL}~\cite{Zorzi2019_3GPPbeamtutorial, Heng2021_BMchallenges}. 

	The existing limited deployments of 5G-NR \ac{FR2} networks~\cite{Verizon_5G, GSMA_Whitepaper} rely on \ac{CBF}, as a low complexity and cost-effective solution, given that \ac{DBF} is prohibitively complex and costly for large antenna arrays requiring a dedicated \ac{RF} chain per antenna element~\cite{SurveyBeamforming_2016}.
	However, \ac{CBF} limits the transmission/reception to a single beam per time slot and \ac{RF} chain over the entire instantaneous bandwidth, \mbox{i. e.} constituting a \ac{TDMA} system~\cite{3GPP_PHYlayer}, which limits the per-UE throughput in \mbox{multi-UE} networks. 
	To better utilize the large bandwidth and directional nature of mm-wave links for beyond-5G deployments, 5G-NR also specifies the use of \ac{HBF} with multiple RF chains to simultaneously serve multiple \ac{UEs} via individual BPLs, \mbox{i. e.} spatial layers in a \ac{MU-MIMO} configuration~\cite{3GPP_PHYlayer}. This improves the spectral efficiency as multiple \ac{UEs} are served with \ac{SDMA} using the same time-frequency resources, and reduces latency as more \ac{UEs} can be served simultaneously instead of waiting for their TDMA time slot. 

	Unlocking the full potential of mm-wave networks by efficiently sharing the network capacity via SDMA comes with many challenges in real network deployments~\cite{Ericsson_Handbook}. 
	HBF \mbox{MU-MIMO} offers an attractive solution for mitigating the co-scheduled UE beam interference via digital precoding for interference cancellation. However, the efficiency of HBF \mbox{MU-MIMO} is predicated on: (\textit{i}) the antenna configuration having sufficient RF chains, (\textit{ii}) the beam codebook having sufficient steering resolution and (\textit{iii}) resource scheduling being dynamic and able to adapt to up-to-date \ac{CSI}. In practice, the predefined gNB/UE codebooks play a key role in acquiring accurate \ac{CSI}. The \mbox{mm-wave} channel can only be observed through the codebook entries which project the received RF signal to the baseband~\cite{hassan_2020_channel}, and should therefore be properly designed to provide sufficient gain and angular steering resolution to estimate the full channel matrix. 	
	Prior mm-wave networking studies on HBF MU-MIMO~\cite{Heath2016_OverviewMIMO, Kularni2016-MIMOComparison, Alkhateeb_2015, Gomez_FullstackHBF_2021, Sun2018, Sun2018_framework, NIF2022, versatileRRM2023} have typically focused on designing the digital precoder for interference cancellation assuming availability of perfect \ac{CSI} at the gNB. In practice, \ac{CSI} reporting is time consuming and subject to the \ac{gNB}/UE codebooks, network control signaling and latency constraints~\cite{Heath2016_OverviewMIMO, Kularni2016-MIMOComparison}, which can lead to system performance degradation of HBF MU-MIMO due to imperfect interference cancellation, \mbox{e. g.}~\cite{Alkhateeb_2015, Sun2018_framework}. Moreover,~\cite{Alizadeh_COMCAS2019} demonstrates that even under ideal digital precoding, there is still a considerable amount of strong residual inter-cell interference, which is highly dependent on the multi-UE BPL allocation for the network. 
A number of mm-wave SDMA studies~\cite{Marzi_2019, BOOST_2020, LiCaireRRM, XueNonOrthBeams2017, Kwon2019_SDMA, Alizadeh_2019, Ichkov_WCNC23} have instead focused on clustering co-scheduled UEs that can be served via spatially separated BPLs for SDMA gains, using CBF with a sufficient number of individual antenna arrays to serve all UEs. However,~\cite{Alizadeh_2019, Ichkov_WCNC23} have shown that under realistic antenna and channel models, \ac{CBF} entails strong residual interference which limits the extent of achievable BPL separation and SDMA gains. Interference management in practical mm-wave SDMA networks should therefore \textit{jointly consider the MU-MIMO interference cancellation and spatial BPL allocation}.  
			
\begin{table*}[t]
		\centering
		\footnotesize
		\resizebox{\textwidth}{!}{ 
		\begin{tabular}{l l l l c c c} \toprule
			{Reference} & {Interference}  &  {Channel} & {Beamforming} &  {Multi-cell} & {Intra-cell} & {Inter-cell}\\ 
			{studies}     & {management} & {model}    & {model}          & {network} & {interference} & {interference}\\ 				\midrule 
	\cite{Heath2016_OverviewMIMO, Kularni2016-MIMOComparison, Alkhateeb_2015, Gomez_FullstackHBF_2021}  	& Interference cancellation  & Geometric  & HBF   & \xmark & \xmark 	 & Added-noise constant	\\
	\cite{Sun2018, Sun2018_framework}  	&   Interference cancellation & 3GPP, NYUSIM & HBF &  \cmark   & \xmark & Estimate \\	
	\cite{NIF2022, versatileRRM2023}  	&   Interference cancellation & Geometric & HBF &  \cmark   & \xmark & Added noise-constant \\				
	\cite{Marzi_2019, BOOST_2020} & Spatial BPL allocation    & 3GPP, NYUSIM & Idealized  &  \cmark & \xmark &  Estimate \\
	\cite{LiCaireRRM, XueNonOrthBeams2017} & Spatial BPL allocation    & Log-distance (LOS only) & Idealized  & \cmark  & \xmark &  Estimate \\
	\cite{Kwon2019_SDMA, Alizadeh_2019}  	& Spatial BPL allocation  & 3GPP, NYUSIM  & \ac{CBF}  & \cmark  & Estimate & Estimate \\
	\cite{Ichkov_WCNC23} & Spatial BPL allocation  & 3D ray-tracing & \ac{CBF}  & \cmark  & \cmark & \cmark \\	
	\textbf{This paper}		&	\textbf{Both}   & \textbf{3D ray-tracing}  & \textbf{HBF} & \cmark  & \cmark & \cmark  \\ 
	\bottomrule 
		\end{tabular}}
\caption{Related work on interference management approaches for mm-wave SDMA networks. \label{tab:related_work}}
\vspace{-10pt}
	\end{table*}

\subsection{Related Work} \label{sec:related_work}

	The prior literature of these two main approaches for interference management in mm-wave SDMA networks  is summarized in Table~\ref{tab:related_work}. The first approach relies on designing a co-scheduled UE precoder for interference cancellation based on \ac{CSI} at the gNB using HBF MU-MIMO~\cite{Heath2016_OverviewMIMO, Kularni2016-MIMOComparison, Alkhateeb_2015, Gomez_FullstackHBF_2021, Sun2018, Sun2018_framework, NIF2022, versatileRRM2023}. The second approach focuses on interference-aware and \mbox{load-balancing} multi-UE BPL allocation using CBF serving UEs via individual antenna arrays~\cite{Marzi_2019, BOOST_2020, LiCaireRRM, XueNonOrthBeams2017, Ichkov_WCNC23, Kwon2019_SDMA, Alizadeh_2019}. 
	
	Interference management for HBF \ac{MU-MIMO}~\cite{Heath2016_OverviewMIMO, Kularni2016-MIMOComparison, Alkhateeb_2015, Gomez_FullstackHBF_2021, Sun2018, Sun2018_framework, NIF2022, versatileRRM2023} typically treats the UE co-scheduling problem within a single cell~\cite{Heath2016_OverviewMIMO, Kularni2016-MIMOComparison, Alkhateeb_2015, Gomez_FullstackHBF_2021}, or limited multi-cell scenarios~\cite{Sun2018, Sun2018_framework, NIF2022, versatileRRM2023}, designing the digital baseband precoder, \mbox{e. g.} \mbox{zero-forcing}, under the assumption of a sufficient number of RF chains and availability of perfect \ac{CSI} at the gNB.	
	This is not always guaranteed in practice, which can lead to performance degradation due to imperfect \ac{CSI} and in turn only partial interference cancellation~\cite{Alkhateeb_2015}. Authors in~\cite{Gomez_FullstackHBF_2021} show the SDMA performance gains of HBF MU-MIMO over CBF and study the end-to-end network implications. However, they only consider a single gNB serving seven UEs, and still show that the network suffers from significant control overhead and delay of acquiring full \mbox{per-UE} CSI. 
	Authors in~\cite{Sun2018, Sun2018_framework, NIF2022, versatileRRM2023} consider HBF \mbox{MU-MIMO} in limited multi-cell network scenarios.~\cite{Sun2018} proposes a HBF MU-MIMO framework based on full CSI and multi-cell cooperation, showing that non-coordinated can outperform fully-coordinated HBF \mbox{MU-MIMO} under availability of only quantized CSI~\cite{Sun2018_framework}. Authors in~\cite{NIF2022} propose a near interference-free UE scheduling (NIF), with~\cite{versatileRRM2023} further generalizing the NIF framework for different SDMA optimization objectives, assuming perfect intra-cell interference cancellation and very limited inter-cell interference. However,~\cite{NIF2022, versatileRRM2023} only assume interference as an added noise constant, and UEs with only \ac{LOS} links to their serving gNB and \ac{NLOS} links to the other interfering gNBs. Importantly,~\cite{Sun2018, Sun2018_framework, NIF2022, versatileRRM2023} assume perfect coordination and synchronization between the gNBs, and neglect the practical overhead and delay constraints of obtaining full per-UE channel estimates, and thus only demonstrate HBF \mbox{MU-MIMO} performance under idealized network~conditions.

	Interference management in \ac{CBF} mm-wave SDMA networks focuses on spatial BPL allocation for clustering co-scheduled UEs that can be sufficiently spatially separated, \mbox{e. g.}~\cite{Marzi_2019, BOOST_2020, LiCaireRRM, XueNonOrthBeams2017, Ichkov_WCNC23, Kwon2019_SDMA, Alizadeh_2019}. Authors in~\cite{Marzi_2019, BOOST_2020, LiCaireRRM} demonstrate SDMA gains when sufficient BPL separation can be maintained to effectively eliminate intra-cell interference, whereas~\cite{XueNonOrthBeams2017} models the interference in the angular domain but considers LOS links only. However, the studies in~\cite{Marzi_2019, BOOST_2020, LiCaireRRM, XueNonOrthBeams2017} rely on statistical channel models and idealized sector antennas, that neglect the spatial distribution of feasible BPLs and the resulting interference implications. Authors in \cite{Kwon2019_SDMA} propose a \ac{CBF} approach using codebooks with low inter-beam overlap, and show that it can outperform HBF \ac{MU-MIMO} assuming quantized \ac{CSI} feedback. However,~\cite{Kwon2019_SDMA} only considers a single gNB serving UEs based on sum-rate clustering and only estimates the interference as added noise. The authors in~\cite{Alizadeh_2019} propose a load-balancing UE association which swaps the worst gNB-UE connection (WCS) at each iteration considering the per-slot interference estimate. In~\cite{Ichkov_WCNC23} we show that under realistic channel considerations, the non-negligible sidelobes of realistic arrays result in significant intra-cell interference which strongly limits the SDMA gains. Additionally,~\cite{Ichkov_WCNC23} proposes an interference-aware BPL allocation for CBF SDMA networks which outperforms the WCS benchmark~\cite{Alizadeh_2019}. In particular,~\cite{Ichkov_WCNC23} demonstrates the benefits of spatial BPL allocation by using secondary BPLs for interference mitigation, going beyond the default \mbox{5G-NR} association to the strongest BPL found during initial access typically considered in the literature, e.g.~\cite{Alkhateeb_2015, Gomez_FullstackHBF_2021, Sun2018, Sun2018_framework, NIF2022, versatileRRM2023, Kwon2019_SDMA, Alizadeh_2019}. 

\subsection{Contributions}
	In this paper, we extend beyond the prior literature \mbox{(\emph{cf.} Table.~\ref{tab:related_work})} and propose an interference-aware BPL allocation (\textit{IABA}) for practical mm-wave HBF MU-MIMO networks that goes beyond the default \mbox{5G-NR} association to the strongest BPL found during initial access. We are thus the first to jointly consider MU-MIMO interference cancellation and spatial BPL allocation for interference management in practical mm-wave HBF MU-MIMO networks. The main contributions of this work are as follows: 
	
	\begin{itemize}
	\item We present an extensive study of multi-cell mm-wave HBF MU-MIMO networks based on site-specific channel data obtained via 3D ray-tracing with realistic building and antenna models, the accuracy of which we have validated against real-world phased antenna arrays measurements~\cite{Raytracing, EuCAP'23}, and network configuration following 3GPP specifications for multi-sector antennas, initial access and channel estimation~\cite{3GPP_model}. 
	
	\item We therefore characterize the residual intra- and inter-cell interference in practical \mbox{mm-wave} HBF MU-MIMO networks, given the spatial distribution of feasible BPLs and \mbox{5G-NR} CSI signaling, showing that interference becomes the main limiting factor of HBF \mbox{MU-MIMO} performance, which is in contrast to prior studies assuming limited to no-interference operation~\cite{Sun2018, Sun2018_framework, NIF2022, versatileRRM2023}.
	
	\item We propose \textit{IABA}, a 5G-NR-compliant BPL allocation, as a solution for mitigating the spatial interference in practical mm-wave HBF MU-MIMO networks. We solve the network sum throughput optimization objective via either a \textit{distributed} or a \textit{centralized} BPL allocation using dedicated \ac{CSI-RS} resources for candidate BPL monitoring. 
	
	\item Our results show that \textit{IABA} offers significant network performance gains over default 5G-NR, by facilitating allocation of secondary BPLs other than the strongest BPL found during initial access. We further demonstrate that under realistic CSI signaling constraints, \textit{IABA} allows \ac{HBF} MU-MIMO to outperform the interference-agnostic \ac{DBF} MU-MIMO baseline. 
	\end{itemize}

\subsection{Organization of the paper}

	The rest of this paper is organized as follows. Sec.~\ref{sec:System_model} details the system model considerations, including channel and multi-panel antenna models, beamforming, channel estimation and multi-UE interference cancellation, and presents key performance evaluation metrics. Sec.~\ref{sec:link_allocation_sec} presents the BPL allocation problem and our \textit{IABA} algorithm and 5G-NR-compliant operation. Sec.~\ref{sec:simulation_model} presents the simulation model, and Sec.~\ref{sec:Results} presents the results. Sec.~\ref{sec:Conclusions} concludes the paper.
	
\section{System Model}
\label{sec:System_model}

\subsection{{mm}-Wave Channel Model}
\label{sec:channel_model}
 
	We model the 3D directional mm-wave channel between the \ac{gNB} and the UE as a set of $K$ LOS and NLOS propagation paths using the geometric channel model~\cite{ayach_2014_spatially}. Assuming a \ac{gNB} as the transmitter with $N_\text{t}$ antenna elements and a UE as the receiver with $N_\text{r}$ antenna elements, we define the geometric $N_\text{r}\times N_\text{t}$ channel matrix $\mathbf{H}$ as 
\begin{equation}\label{eq:analytical_channel}
\mathbf{H} = \sqrt{\frac{N_\text{r}N_\text{t}}{K}} \sum_{k=1}^{K} \alpha_{k} \mathbf{a}_{\mathrm{r}}\left(\phi_{k}^\text{r}, \theta_{k}^\text{r}\right) \mathbf{a}_{\mathrm{t}}^{H}\left(\phi_{k}^\text{t}, \theta_{k}^\text{t}\right),
\end{equation}
\noindent where $\alpha_k$ denotes the complex channel gain of the $k^\text{th}$ path, and $\{\phi_{k}^{\mathrm{t}}, \theta_{k}^{\mathrm{t}}\}$ and $\{\phi_{k}^{\mathrm{r}}, \theta_{k}^{\mathrm{r}}\}$ denote the angle of departure (AoD) and angle of arrival (AoA) in azimuth and elevation, respectively. 
The transmit and receive antennas are described by their steering vectors $\mathbf{a}_{\mathrm{t}}$ and $\mathbf{a}_{\mathrm{r}}$, which are identical to the phase-shifting weights applied to steer the antenna arrays towards the AoD and AoA, respectively~\cite{hassan_2020_channel}.	

	For the special case of an $N$ element uniform linear array (ULA) positioned along the $y$-axis and steered towards angle $\phi$ in azimuth, the array steering vector is given as
\begin{equation}\label{eq:response_ULA}
\resizebox{.91\hsize}{!}{$
\mathbf{a}_{\mathrm{ULA}}(\phi)=\frac{1}{\sqrt{N}}\left[1\; e^{j 2 \pi \frac{d}{\lambda} \sin (\phi)} \cdots e^{j(N-1) 2 \pi \frac{d}{\lambda} \sin (\phi)}\right]^{T}.
$}
\end{equation}
	
\noindent We generalize it to a \ac{URA} in the \mbox{$y$-$z$~plane}, with the array steering response given as
\begin{equation}\label{eq:response_URA}
	\mathbf{a}_{\text{URA}}(\theta, \phi)=\mathrm{a}_{\text{h, ULA}}(\phi) \otimes \mathrm{a}_{\text{v, ULA}}(\theta),
\end{equation}
	where $\otimes$ denotes the Kronecker product of the horizontal steering vector determining the azimuth response $\mathrm{a}_{\text{h, ULA}}(\phi)$ and the vertical steering vector for the elevation response $\mathrm{a}_{\text{v, ULA}}(\theta)$ of a same sized ULA~(\ref{eq:response_ULA}).

	Finally, we express the downlink received signal $\mathbf{y}$ as 
\begin{equation}\label{eq:RXsignal}
	\mathbf{y} = \mathbf{w}_\text{c}^\text{H} \mathbf{H} \mathbf{W}_\text{p} \mathbf{s} + \mathbf{w}_\text{c}^\text{H}\mathbf{n},
\end{equation}
	where $\mathbf{w}_{\text{c}}^\text{H}$ denotes the Hermitian of the combining vector, $\mathbf{W}_{\text{p}}$ denotes the precoding matrix, $\mathbf{s}$ denotes the data stream vector, and $\mathbf{n}$ denotes the noise vector.

	We use an open-source mm-wave ray-tracing tool to obtain site-specific directional channel data based on real 3D building and antenna models~\cite{Raytracing, EuCAP'23}. We perform dedicated ray-tracing runs for each gNB position in our multi-cell network (\emph{cf.}~Sec.~\ref{sec:simulation_model}), considering free-space propagation and strong reflections (up to two-bounces) as dominant propagation mechanisms; diffraction is neglected due to its limited role at mm-wave~\cite{Reflection_28GHz}. The \mbox{ray-tracing} output is used to reconstruct the channel matrix as per~(\ref{eq:analytical_channel}). 
	
\begin{figure}[tb]
\centering
\includegraphics[width=0.4\textwidth]{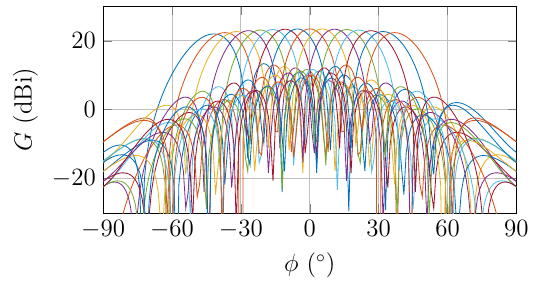}
\vspace{-5pt}
\caption{Example codebook with 16 beams for a sector antenna covering the steering range of $\phi \in (-45^\circ,~45^\circ)$.}
\label{fig:codebook}
\vspace{-10pt}
\end{figure}		

\begin{figure*}[t]
\centering
\includegraphics[width=0.99\textwidth]{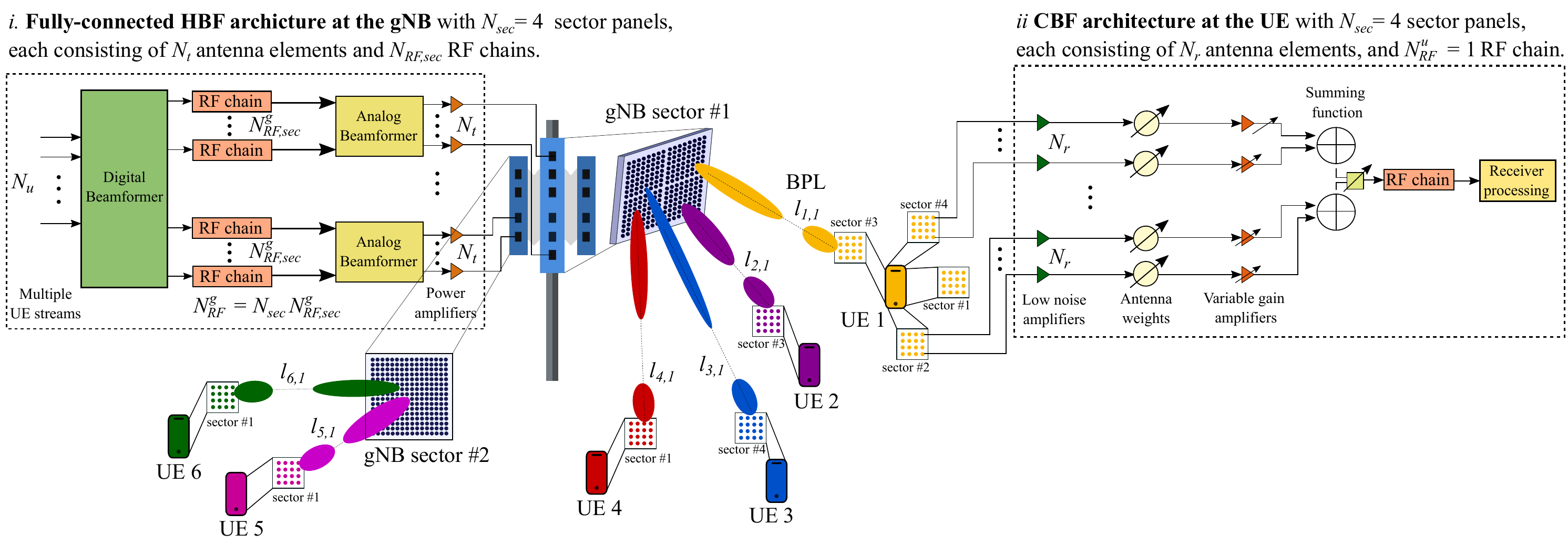} 
\vspace{-5pt}
\caption{Illustration of a mm-wave network, with an example scenario of a gNB simultaneously serving $N_u=6$~UEs. The gNB is using a fully-connected HBF architecture with $N_{sec}=4$ sector panels each equipped with $N_t$ antenna elements and $N^g_{RF,sec}=4$ RF chains, i. e. the gNB can simultaneously serve up to $N^g_{RF}=16$~UEs . For clarity, we only illustrate two out of the four gNB sector panels, with four UEs being served by sector \#$1$ and two UEs served by sector \#$2$. The UEs use a CBF architecture with $N_{sec}=4$ sector panels, each equipped with $N_r$ antenna elements, and $N^u_{RF}=1$ RF chain.}
\label{fig:1}
\vspace{-10pt}
\end{figure*}

\subsection{Multi-Panel Antenna \& Beamforming Model}
\label{sec:antenna_BPL_model}

	For the transmit/receive antennas, we employ the multi-panel antenna model~\cite{3GPP_model}, where the gNB/UE is equipped with \mbox{$N_{sec}=4$} sector antenna panels, each covering one sector/quadrant of the angular space. We assume that the gNB/UE has unified control over all its multi-panel antennas, which have a fixed initial orientation. To steer the individual beams, we assume predefined codebooks for each sector antenna panel, $\mathcal{C}^{g}_{sec}$ at the gNB and $\mathcal{C}^{u}_{sec}$ at the UE of size $n_q$ bits, which consist of $2^{n_q}$ beams covering the angular steering range of $\pm45^\circ$, following 3GPP's specifications in \mbox{Table.~7.3-1} in~\cite{3GPP_model}. An example of such a sector antenna codebook consisting of 16 beams is shown in Fig.~\ref{fig:codebook}.
	
	For the beamforming, we assume a fully-connected \ac{HBF} architecture at the gNB, mapping all RF chain outputs per sector panel onto a common analog beamformer, which applies a sum of all phase-shifting outputs to each of the $N_t$ elements of the sector antenna panel, as shown in Fig.~\ref{fig:1}. Each RF chain controls a part of the overall sector antenna panel, with a total of $N^g_{RF,sec}$ RF chains per panel. Thus, the maximum number of UEs that can be served simultaneously at the gNB is constrained by the total number of RF chains , i. e. $N_{u} \leq N^g_{RF}$, where $N^g_{RF}=N_{sec}N^g_{RF, sec}$. 
	At the UE, we assume a \ac{CBF} architecture following 5G-NR specifications in~\cite{3GPP_model}, such that each UE is equipped with $N^u_{RF}=1$ RF chain controlling $N_{sec}=4$ sector antenna panels, each consisting of $N_r$ antenna elements, as shown in Fig.~\ref{fig:1}. 

\subsection{Initial Access: Initial Serving BPL Selection}
\label{sec:initial_access}

	Steering the antenna beams to the correct orientation for establishing a mm-wave \ac{BPL} necessitates beam training~\cite{Zorzi2019_3GPPbeamtutorial}. A mm-wave \ac{BPL} $l_{i,j}$ is defined by the UE~$i$, the serving gNB $j$, and the respective steering orientations at the gNB $\{\phi_{l_{i,j}}^{j}, \theta_{l_{i,j}}^{j}\}$ and the UE $\{\phi_{l_{i,j}}^{i}, \theta_{l_{i,j}}^{i}\}$. 
	 We follow the 5G-NR initial access as our beam training procedure~\cite{Zorzi2019_3GPPbeamtutorial}, and assume downlink control operation based on an exhaustive beam sweep over the codebook entries of $\mathcal{C}^{j}$ at gNB~$j$ and $\mathcal{C}^{i}$ at UE~$i$, where 
	 $\mbox{$\mathcal{C}^{j}=\bigcup\limits_{sec=1}^{N_{sec}} \mathcal{C}^{j}_{sec}$}$ and $\mbox{$\mathcal{C}^{i}=\bigcup\limits_{sec=1}^{N_{sec}} \mathcal{C}^{i}_{sec}$}$ 
	 denote the complete gNB and UE codebooks consisting of all the entries for each sector antenna codebook.	 
	 The initial serving BPL $l_{i,j}$ for UE~$i$ is then selected out of the finite set of feasible BPLs found during initial access over all gNBs,
$\mbox{$\mathbf{N}_{l,i}=\bigcup\limits_{j \in \mathbf{G}} \bigcup\limits_{n=1}^{|\mathcal{C}^{j}|} \bigcup\limits_{m=1}^{|\mathcal{C}^{i}|} {l}_{{i,j,n,m}}$}$
as the one which maximizes the \ac{SSB} \ac{RSRP}, 
and is determined as 
	 \begin{equation} \label{eq:initialBPL}
	 l_{i, j} ~  \leftarrow ~ \underset{\substack{l_{i, j} \in \mathbf{N}_{l,i} \\ \forall \mathbf{w}_{\text{p},l_{i,j}} \in \mathcal{C}^j \\ \forall \mathbf{w}_{\text{c},l_{i,j}} \in \mathcal{C}^i }}{\operatorname{argmax}} ~ P_{SSB} \left\| \mathbf{w}_{\text{c},l_{i,j}}^\text{H} \mathbf{H}_{i, j} \mathbf{w}_{\text{p},l_{i,j}} \right\|^2,~~\forall j \in \mathbf{G}
	 \end{equation}
	
\noindent where $P_{SSB}$ denotes the SSB transmit power, $\mathbf{G}$ denotes the set of gNBs in the network, and $\mathbf{w}_{\text{p},l_{i,j}}$ and $\mathbf{w}_{\text{c},l_{i,j}}$ denote the precoding and combining vectors applied to steer the arrays towards the AoD and AoA of the BPL $l_{i,j}$, respectively.

\subsection{Channel Estimation}
\label{sec:channel_estimation}

	The exhaustive beam sweep of the finite gNB/UE codebooks determines the steering orientations of the candidate BPLs that best correspond to the real AoD/AoA of the $K$ LOS/NLOS channel propagation paths in~(\ref{eq:analytical_channel}), within the margins of the codebook steering resolution. To accommodate for the inaccurate angular information during channel estimation, we use the wide-angle steering codebook which generates variable beamwidth patterns to fit a desired steering resolution for a given antenna array~\cite{peng_2018_robust}, i. e. resulting in quantized channel estimates due to use of codebooks with finite steering resolution. Specifically, for both the gNB and the UE, we may define an azimuth steering codebook $\mathcal{C}_\phi$ of size $n_q$ bits, which is able to distinguish between $2^{n_q}$ angles over the range of $90^\circ$ for each of the $N_{sec}=4$ antennas. Similarly, we may define an elevation steering codebook $\mathcal{C}_\theta$ of size \mbox{$n_q-1$} bits over the range of $180^\circ$ for each of the $N_{sec}=4$ antennas. 
	 The finite channel estimation resolution over the azimuth/elevation steering ranges is thus given as 
\begin{equation}\label{eq:ch_est_resolution}
\phi_\mathrm{res} = \frac{360^\circ}{N_{sec}\cdot 2^{n_q}},~~~
\theta_\mathrm{res} = \frac{180^\circ}{N_{sec}\cdot 2^{n_q-1}}. 
\end{equation}	
	The quantized angle estimates of the real AoD for the channel propagation paths in~(\ref{eq:analytical_channel}) are defined as 
\begin{equation} \label{eq:angle_quant1}
\hat{\phi}_{i,k} = \underset{\phi\in \mathcal{C}^i_\phi}{\operatorname{argmin}} \left\| \phi -  \phi^t_{k}\right\|,~~
\hat{\theta}_{i,k} = \underset{\theta\in \mathcal{C}^i_\theta}{\operatorname{argmin}} \left\| \theta -  \theta^t_{k}\right\|,
\end{equation}
	where the same principle holds equivalently for the AoD/AoA and each channel propagation path $k$ in~(\ref{eq:analytical_channel}). 	
	We note that the use of finite resolution codebooks results in an inability to distinguish individual channel paths with the same quantized angles. In such cases, the channel estimation algorithm reports a single path with quantized parameters, \mbox{i. e.} the individual paths are merged into one known path and their complex gains are combined.
	Finally, the estimated channel matrix $\hat{\text{H}}_{i}$ for UE~$i$ is given as
\begin{equation} \label{eq:quant_channel}
\hat{\text{H}}_{i} = {\begin{pmatrix}
\hat{\text{H}}_{1,1} & \cdots  & \hat{\text{H}}_{1,4} \\
\vdots  & \ddots  & \vdots  \\
\hat{\text{H}}_{4,1} & \cdots  & \hat{\text{H}}_{4,4} \\
\end{pmatrix}},
\end{equation}
where \mbox{e. g.} $\hat{\text{H}}_{1,4} \in \mathbb{C}^{N_\text{r} \times N_\text{t}}$ denotes the estimated channel matrix between the first sector antenna panel of UE~$i$ and the fourth sector antenna panel at the serving gNB $j$. 

	Following 3GPP's specifications, we distinguish between gNB/UE codebooks used for channel estimation and initial access, as well as reference signals used for each procedure~\cite{3GPP_NR}.
	Assuming downlink operation using the same codebook for channel estimation and initial access, the number of reference signals for CSI reporting is limited to $64$, in line with 3GPP specifications for the number of SSBs at FR2~\cite{Zorzi2019_3GPPbeamtutorial}. In 5G-NR HBF MU-MIMO terms~\cite{3GPP_NR}, this corresponds to CSI reporting for Type-II codebooks with no oversampling and a sector antenna codebook size of $n_q=4$ bits, i. e. 16 beams per each of the four sector antennas and 64 beams in total. Type-II codebooks also support a more detailed CSI-RS-based reporting~\cite{Qin2023_CSIcodebook}, with an oversampling factor of up to four for both azimuth and elevation, making a codebook size of $n_q=6$ bits also feasible, i. e. 64 beams per each of the four sector antennas and a total of 256 beams. 
	
	In Sec.~\ref{sec:Results} we vary the size of the gNB/UE codebooks used for channel estimation, and study the corresponding accuracy of the estimated channel matrix which is then used for designing the digital baseband precoder for multi-UE interference cancellation~(\emph{cf.}~Sec.~\ref{sec:multiUE_IC}). We additionally consider as a reference the ideal case, where the channel estimates are acquired using an infinite resolution codebook $n_q=\infty$, such that the channel path angles and complex gains are assumed to be perfectly known, \mbox{i. e.} $\hat{\text{H}}=\text{H}$.

\subsection{Multi-User Interference Cancellation}
\label{sec:multiUE_IC}

	 The computation of the precoding weights for serving the multiple UEs at the gNB is split into two stages, as per~\cite{Alkhateeb_2015}.
	In the first stage, following the initial exhaustive beam sweep, the codebook entries which maximize the RSRP are selected to steer the arrays at the gNB $\mathbf{w}_{\text{p, RF},i}$, and the UE $\mathbf{w}_{\text{c},i}$, towards the AoD/AoA of the serving BPL $l_{i,j}$~(\ref{eq:initialBPL}). 
	
	In the second stage, we need to compute the effective channel for each UE~$i$ based on the estimated channel $\mathbf{\hat{H}}_i$~(\ref{eq:quant_channel}). For this, we use the precoding $\mathbf{w}_{\text{p, RF},i}$ and combining $\mathbf{w}_{\text{c},i}$ vectors from the first stage, as  
	\begin{equation}\label{eq:effective_channel}
	\mathbf{\hat{H}}_{\text{eff},i} = \mathbf{w}_{\text{c},i}^{\text{H}} \mathbf{\hat{H}}_i \mathbf{W}_\text{p,RF},
	\end{equation}
	where $\mathbf{W}_\text{p,RF} \leftarrow \left[\mathbf{w}_\text{p,RF,1} \cdots \mathbf{w}_{\text{p,RF},N_u} \right]$ denotes the aggregate RF precoding matrix for the $N_u$ UEs served by the gNB. Each UE feeds this channel information back to the gNB, with the effective channel vector reflecting a drastically reduced dimensionality, i. e. $N_u  \ll N_{sec} N_t$. We note that the accuracy of the effective channel depends on the estimated channel matrix $\mathbf{\hat{H}}_i$, which is dictated by the gNB/UE codebooks used for channel estimation (\emph{cf.}~Sec.~\ref{sec:channel_estimation}).
	
	The gNB then designs the digital baseband precoder based on the UE channel feedback using zero-forcing for multi-UE interference cancellation as
	\begin{equation}\label{eq:zero_forcing}
	\mathbf{W}_{\text{p,BB}} = \mathbf{\bar{H}} \left(\mathbf{\bar{H}}^\text{H} \mathbf{\bar{H}}\right)^{-1},
	\end{equation}	
\noindent where $\mathbf{\bar{H}} = \left[\mathbf{\hat{H}}_{\text{eff},1}^\text{H} \cdots \mathbf{\hat{H}}_{\text{eff},N_u}^\text{H}\right]$ represents the aggregated effective channels for the $N_u$ UEs.
	The resulting baseband precoding vectors for each UE~$i$ are then normalized to fulfill the power constraint,
	$\mathbf{w}_{\text{p,BB},i} =  \frac{\mathbf{w}_{\text{p,BB},i}}{\| \mathbf{W}_\text{p,RF} \mathbf{w}_{\text{p,BB},i} \|_F}$, 
	where $\|~\|_F$ denotes the Frobenius norm. 
	
\noindent Finally, the individual precoding vector $\mathbf{w}_{\text{p},i}$ for UE~$i$ in (\ref{eq:RXsignal}) is generated by multiplying the RF and baseband~precoding,
	\begin{equation}\label{eg:precoding}
	\mathbf{w}_{\text{p},i} = \mathbf{W}_\text{p,RF} \mathbf{w}_{\text{p,BB,i}},
	\end{equation}
\noindent with $\mathbf{W}_\text{p,RF} \in \mathbb{C}^{N_{sec} N_t\times N_u}$ and $\mathbf{w}_{\text{p,BB,i}} \in \mathbb{C}^{N_u \times N_u}$. 

	Effectively, the gNB precoder in the case of HBF maps the $N_u$ UE streams to the $N^g_{RF}$ RF chains, thus significantly reducing the complexity compared to \ac{DBF}.

\subsection{Performance Evaluation Metrics}
\label{sec:evaluation_metrics}
	 
	The \ac{SINR} for UE~$i$ served by gNB~$j$ via the serving BPL $l_{i,j}$ is given as 
	\begin{equation} \label{eq:SINR}
	SINR_{i} =  \frac{RSS_{i}} {I_{intra,i} + I_{inter, i} + N},
	\end{equation}
\noindent where $N$ denotes the noise power, consisting of thermal noise over the system bandwidth plus a noise figure, and the \ac{RSS} is given by
	\begin{equation} \label{eq:RSS}
	RSS_i = P_\text{i} \| \mathbf{w}_{\text{c},i}^\text{H} \mathbf{{H}}_{i,j} \mathbf{w}_{\text{p},i}\|^2,
	\end{equation}
\noindent where $P_\text{i}$ denotes the gNB transmit power allocated to UE~$i$ with the total transmit power $P_{max}$ equally shared among all allocated UEs at the gNB, and the precoding $\mathbf{w}_{\text{p},i}$ and combining $\mathbf{w}_{\text{c},i}$ vectors determined as in Sec.~\ref{sec:multiUE_IC}.

	We define the intra-cell interference $I_{intra, i}$ and inter-cell interference $I_{inter, i}$ for UE~$i$ served by gNB~$j$ as 
	\begin{equation} \label{eq:intra-cell}
	I_{\text{intra},i} = P_\text{U} \| \mathbf{w}_{\text{c},i}^\text{H} \mathbf{{H}}_{i,j} \mathbf{W}_\text{p}^j \|^2,
	\end{equation}
	\begin{equation} \label{eq:inter-cell}
	I_{\text{inter},i} = P_\text{U} \| \mathbf{w}_{\text{c},i} \sum_{g\neq j} \mathbf{{H}}_{i,g} \mathbf{W}_\text{p}^g \|^2,
	\end{equation}
\noindent where $P_\text{U}$ denotes the gNB transmit power allocated to each UE and \mbox{$\mathbf{W}_\text{p}^j = \left[\mathbf{w}_{\text{p},1}, \cdots, \mathbf{w}_{\text{p},N_u}\right]$} denotes the precoding matrix for the $N_u$ UEs served by the gNB.
	
	Lastly, to map the $SINR_{i}$ to the achievable throughput $R_{i}$, we adopt the wideband 5G-NR model~\cite{Verizon_5G} using an attenuated, truncated Shannon bound as 
\begin{eqnarray}
\small
\label{eq:rate}
R_{i} = \begin{cases} 
0 &  SINR_{i} < SINR_{min}, \\
R_{max} &  SINR_{i} \geq SINR_{max}, \\ 
\alpha B \log_2 (1+SINR_{i}) & \mbox{otherwise},
\end{cases}
\end{eqnarray}
\noindent where $\alpha=0.75$ represents the implementation losses,  \mbox{$B=400$~MHz} denotes the bandwidth, \mbox{${SINR}_{min}=-5$}~dB denotes the coverage threshold and \mbox{${SINR}_{max}=20.05$}~dB is the SINR required to get the maximum achievable throughput of $R_{max}=2$~Gbps.

\section{Multi-User BPL Allocation in HBF \mbox{MU-MIMO} mm-Wave Networks}
\label{sec:link_allocation_sec}

\subsection{BPL Allocation Problem}
\label{sec:link_problem}

	Assuming that a UE is only served by a single gNB via one BPL, we formulate the multi-UE BPL allocation that maximizes the network sum throughput as
	\begin{equation} \label{eq:alloc}
	\mathbf{L} = \argmax_{\mathbf{L} \subset  \mathbf{N}_l} \; \sum\limits_{i \in \mathbf{U}} R_{i}, 
	\end{equation}
\begin{subequations}
	\begin{equation} \label{eq:coverageConstraint}
	\text{s.t.}~SINR_{i} \geq SINR_{min},  \forall i \in \mathbf{L}
	\end{equation}
	\begin{equation} \label{eq:allocationConstraint}
	|\mathbf{L}_j| \leq N^j_{RF},  \forall j \in \mathbf{G}
	\end{equation}
	\begin{equation} \label{eq:powerConstraint}
	~\sum\limits_{i \in \mathbf{L}_j} P_{i} \leq P_{max},  \forall j \in \mathbf{G}
	\end{equation}
\end{subequations}

\noindent where $\mathbf{L}$ denotes the set of allocated BPLs and $\mbox{$\mathbf{N}_{l}=\bigcup\limits_{i=1}^{N_{UE}} \mathbf{N}_{{l,i}}$}$ denotes the set of candidate BPLs for all UEs. The constraint in (\ref{eq:coverageConstraint}) denotes the coverage threshold for BPL allocation. The constraint in (\ref{eq:allocationConstraint}) denotes the maximum number of UEs that can be served simultaneously on each gNB. The constraint in (\ref{eq:powerConstraint}) denotes the gNB transmit power allocation, with the total power equally split among the served UEs. We note that the combinatorial allocation problem~(\ref{eq:alloc}) can be solved by exhaustive search, but this is computationally infeasible for all but very small networks.

\subsection{IABA: Interference-Aware BPL Allocation}
\label{sec:link_allocation}
	
	We propose \textit{IABA}, an interference-aware BPL allocation scheme which aims to mitigate spatial interference in HBF \mbox{MU-MIMO} mm-wave networks and maximize the network sum throughput~(\ref{eq:alloc}). \textit{IABA} is fully 5G-NR-compliant, and relies on monitoring candidate BPLs via a centralized \ac{CSI-RS} allocation, i. e. \textit{cIABA}, or a distributed, per-gNB \ac{CSI-RS} allocation, i. e. \textit{dIABA}. The detailed step-by-step operation of \textit{IABA} is as follows (\emph{cf.}~Alg.~\ref{alg:JointBPL}). 		

		\begin{algorithm}[t!]
			\caption{Interference-Aware BPL Allocation (\textbf{IABA})}
			\label{alg:JointBPL}
			\small
			\renewcommand{\algorithmicrequire}{\textbf{Initialize:}}
			\renewcommand{\algorithmicensure}{\textbf{Output:}}
			\begin{algorithmic} [1]
				\Require {$\mathbf{U}$, $\mathbf{G}$, \textbf{L}, $\mathcal{C}^{g}$, $\mathcal{C}^{u}$, $N^g_{RF}$, $N^u_{RF}$, $N_{CSI-RS}$.} 
				\State \textbf{Step 1. Initial user association}
				\For {each UE $i$ in $\mathbf{U}$} 
						 \State 	$\mbox{$\mbox{$\mathbf{N}_{l,i}=\bigcup\limits_{j \in \mathbf{G}} \bigcup\limits_{n=1}^{|\mathcal{C}^{j}|} \bigcup\limits_{m=1}^{|\mathcal{C}^{i}|} {l}_{{i,j,n,m}}$}$}$, set of all candidate BPLs. 
						 \State $l_{i,c} ~ \leftarrow ~ \underset{l_{i, c} \in \mathbf{N}_{l,i}}{\operatorname{argmax}}~RSS_{i}$, initial serving BPL on gNB~$c$~(\ref{eq:initialBPL}). 
				\EndFor
				\State \textbf{Step 2. CSI-RS candidate BPL monitoring}
				\For {each UE $i$ in $\mathbf{U}$}   
				\If{~(\textit{distributed IABA})}
					\State $\textbf{C}_{l,i}\leftarrow \underset{RSS}{\text{sort}}~(\mathbf{N}_{l,i,c})$, sorted set of candidate BPLs on the initial serving gNB~$c$.
					\State $|\textbf{C}_{l,i}| = N_{CSI-RS}$, limit number of monitored BPLs.
				\Else{~(\textit{centralized IABA})}
					\State $\textbf{C}_{l,i} \leftarrow \underset{RSS}{\text{sort}}~(\mathbf{N}_{l,i})$, sorted set of all candidate BPLs.
					\State $|\textbf{C}_{l,i}| = N_{CSI-RS}$, limit number of monitored BPLs.
				\EndIf
				\EndFor
				\State \textbf{Step 3. Multi-UE BPL allocation}  
				\For {each UE~$i$ in $\mathbf{U}$} 
					\State	  $\mathbf{C}_{SINR_i}$ = \{\}, provisional BPL allocation set for UE~$i$.
					\For {each $l_{i,j}$ in $\textbf{C}_{l,i}$}
							\State \textit{Compute precoding/combining weights (\ref{eq:effective_channel})--(\ref{eg:precoding}):} 
							\State $\{\mathbf{w}_{\text{p,}k}\}_{k=1}^{|\mathbf{L}_{j}|+1}, \{\mathbf{w}_{\text{c,}k}\}_{k=1}^{|\mathbf{L}_{j}|+1}$, $\forall k \in \{\textbf{L}, l_{i,j}\}$.
							\State \textit{Update metrics of allocated BPLs~(\ref{eq:SINR}):} 
							\If{(\textit{distributed IABA})}
							\State~$SINR_k$, $\forall k \in \{\textbf{L}_{j}, l_{i,j}\}$. 
							\Else{~(\textit{centralized IABA})}
							\State~$SINR_k$, $\forall k \in \{\textbf{L}, l_{i,j}\}$.
							\EndIf
							\State \textit{Provisional BPL allocation:} 
							\If {$(SINR_k \geq SINR_{min}$ \& $|\mathbf{L}_{j}| \leq N^j_{RF}$), $\forall k \in \textbf{L}$}
							\State $\mathbf{C}_{{SINR_i}} = \{ \mathbf{C}_{{SINR_i}}, l_{i,j} \}$. 
							\Else 
							\State \textbf{continue}
							\EndIf
					\EndFor
										
					\If {$\textbf{empty}(\mathbf{C}_{{SINR_i}})$}
					\State \textit{drop UE $i$ from network.}
					\Else 
					\State	 $l_{i,j} \leftarrow \underset{ l_{i,j} \in \mathbf{C}_{SINR_i}}{\operatorname{argmax}} SINR_{i}$, ~
					$\mathbf{L} \leftarrow l_{i, j}$.
					\EndIf
				\EndFor
				\Ensure  {\textbf{L}, final BPL allocation set.}
			\end{algorithmic}
		\end{algorithm}
			
	In the first step, the set of candidate BPLs $\mathbf{N}_{l, i}$ for each UE~$i$ is established following the exhaustive beam sweep of the gNB/UE codebooks during initial access. The UE is initially associated to the gNB which provides the strongest BPL as per~(\ref{eq:initialBPL}).
	In the second step, the network sets up the monitoring of candidate BPLs by allocating dedicated CSI-RS resources. Each UE can monitor up to $N_{CSI-RS}$ BPLs, either solely on the serving gNB using \textit{dIABA} or over the complete network using \textit{cIABA}.
	In the third step, the multi-UE BPL allocation is executed by selecting among the $N_{CSI-RS}$ monitored BPLs using CSI-RS measurement reporting. The precoding/combining weights are computed for each new candidate BPL using (\ref{eq:effective_channel})--(\ref{eg:precoding}), taking into account all previously allocated BPLs on the serving gNB. 
Following this, the network updates the $SINR$ of the co-allocated BPLs, i. e. solely on the serving gNB for \textit{dIABA} or across all gNBs for \textit{cIABA}. The decision on whether the candidate BPL is allocated is then made in an interference-aware manner, as follows.
If none of the previously allocated BPLs are degraded below the coverage threshold~(\ref{eq:coverageConstraint}), and the gNB has enough physical resources to serve all allocated UEs~(\ref{eq:allocationConstraint}), the candidate BPL is provisionally allocated to~$\mathbf{C}_{{SINR_i}}$. Otherwise, the candidate BPL is discarded. Once the $N_{CSI-RS}$ candidate BPLs for the UE have been checked, the candidate BPL in~$\mathbf{C}_{{SINR_i}}$ providing the highest $SINR$ is allocated to the final BPL allocation set~$\mathbf{L}$. If none of the candidate BPLs are eligible for allocation, the UE is dropped.

	In Sec.~\ref{sec:Results}, we evaluate the proposed \textit{IABA} against the default interference-agnostic 5G-NR baseline, where each UE is allocated its strongest BPL as found during initial access (Step 1 of Alg.~\ref{alg:JointBPL}), such that new allocations can lead to previously allocated BPLs being dropped from the network due to excessive interference. 

	\textbf{Complexity Analysis}: Considering closely the multi-UE BPL allocation in Step~3 in Alg.~\ref{alg:JointBPL}, line 21/23 have the highest cost due to the computation of the updated $SINR$ for all co-allocated BPLs. This involves the multiplication of the updated combining, channel, and precoding matrices as per~(\ref{eq:SINR}), which is of complexity order $\mathcal{O} (N_t^2)$~\cite{Heath2016_OverviewMIMO}. We note that the selection of the serving BPL for each UE $i$ is then done in line 32 in Alg.~\ref{alg:JointBPL} by finding the maximum SINR BPL in the provisional allocation set $\mathbf{C}_{{SINR_i}}$, which requires $|\mathbf{C}_{{SINR_i}}|\text{log}(|\mathbf{C}_{{SINR_i}}|)$ computations.
	
	\textit{dIABA} runs independently on each gNB, having \mbox{$|\mathbf{C}_{l,i}|N_{UE}/N_{gNB}$} iterations on average, where $N_{UE}=|\mathbf{U}|$ and $N_{gNB}=|\mathbf{G}|$. In each iteration, the candidate BPLs for each associated UE are checked, in total \mbox{$|\mathbf{C}_{l,i}| \leq K$}, where $K$ denotes the number of channel propagation paths between UE~$i$ and its serving gNB~$j$ in~(\ref{eq:analytical_channel}). We note that the exact number of candidate BPLs between a UE and its serving BS depends on their relative positions, site-specific channel and codebook considerations. The computational complexity of \textit{dIABA} is thus given as $\mathcal{O}\left(K  \frac{N_{UE}^2}{N_{gNB}}  N_t^2 \right).$

	\textit{cIABA} runs a single loop with $N_\text{UE}$ iterations, considering potentially all candidate BPLs over the complete network, which constitutes $ |\mathbf{C}_{l,i}| N_{gNB} \leq K N_{gNB}$ repetitions of SINR calculations. The computational complexity of \textit{cIABA} is thus given as~$\mathcal{O}\left(K  N_{gNB} N_{UE}^2  N_t^2\right).$ 

This shows that with network densification, the computational complexity of the centralized allocation \textit{cIABA} scales with $\mathcal{O} (N_{UE}^2 N_{gNB})$, whereas for \textit{dIABA} it scales with $\mathcal{O} (N_{UE}^2 / N_{gNB})$.

	For the default 5G-NR BPL allocation, the calculation for each UE involves $N_{UE}/N_{gNB}$ calculations on the serving gNB, as we need to calculate the SINR for all allocated BPLs including the new one, but only for the per-UE best RSS BPL found during initial access; this is done for every gNB so in total $N_{gNB}$ iterations. Thus, the computational complexity of default 5G-NR is given as $\mathcal{O}\left(\frac{N_{UE}^2}{N_{gNB}} N_t^2 \right).$

\subsection{5G-NR-Compliant IABA Operation}
\label{sec:5GNR_allocation}

\begin{figure}[t!]
	\centering
	\includegraphics[width=0.49\textwidth]{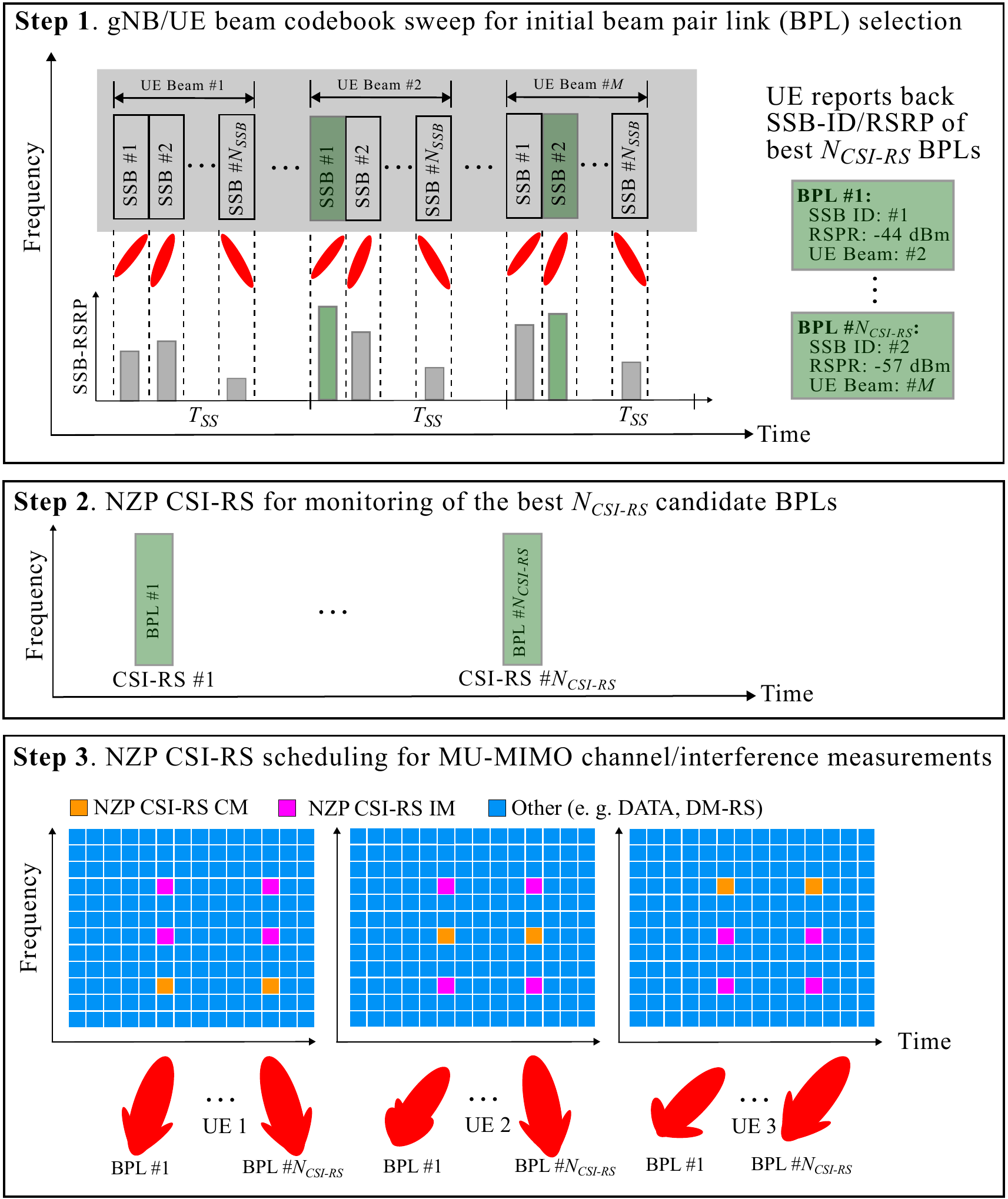}
	\caption{5G-NR-compliant operation for \textit{IABA}: (\textit{i}) SSB-based initial beam sweep for BPL selection, (\textit{ii}) scheduling of NZP CSI-RS for BPL monitoring, (\textit{iii}) NZP CSI-RS scheduling for MU-MIMO channel/interference measurements.}
   	 \label{fig:SSBCSIRS}
   	 \vspace{-5pt}
\end{figure}

	The proposed \textit{IABA} is fully compliant with the 5G-NR standard, and relies solely on standardized reference signals for dedicated channel measurements (CM) and interference measurements (IM)~\cite{Eko2018_CSIreporting}. The detailed step-by-step 5G-NR standard-compliant operation of \textit{IABA} is illustrated in Fig.~\ref{fig:SSBCSIRS}.

	For initial access, we follow the downlink control operation where the gNBs transmit SSB reference signals, such that each SSB corresponds to a different beam index (ID) in the gNB codebook (\emph{cf.}~Step~1 in Fig.~\ref{fig:SSBCSIRS}). Following 3GPP's specifications, the SSBs are allocated in the first 5~ms of an SS burst, which is transmitted with a periodicity of $T_{SS}=20$~ms. We note that 3GPP limits the maximum number of SSBs $N_{SSB}$ up to 64 for FR2~\cite{Zorzi2019_3GPPbeamtutorial}.
	During the transmission of an SS burst, the UE keeps its beam fixed, and continues to sweep all of its beams over consecutive SS bursts, as illustrated in Fig.~\ref{fig:SSBCSIRS}. A complete beam sweep of the gNB/UE codebooks thus takes $MT_{SS}$, where $M=|\mathcal{C}^{u}|$ denotes the number of beam entries in the UE codebook. The UE sends back a measurement report of the beam sweep, including the SSB-RSRP, SSB-ID and UE beam ID of the best $N_{CSI-RS}$ candidate BPLs, as shown in Fig.~\ref{fig:SSBCSIRS}. The UE is consequently initially associated to the gNB which provides the strongest BPL in terms of SSB-RSRP~(\emph{cf.}~Sec.~\ref{sec:initial_access}).
	
	Following the initial access, the network sets up the monitoring of candidate BPLs (\emph{cf.}~Step.~2 in Fig.~\ref{fig:SSBCSIRS}). For this, \textit{IABA} uses dedicated non-zero power (NZP) CSI-RS resources to monitor the best $N_{CSI-RS}$ candidate BPLs found during initial access for each UE, with $N_{CSI-RS}\leq4$ for FR2~\cite{Zorzi2019_3GPPbeamtutorial}. After selecting the per-UE $N_{CSI-RS}$ BPLs for active monitoring, the network schedules the NZP \mbox{CSI-RS} CM/IM resources following the \mbox{5G-NR} standardized methodology~\cite{5GNR_IM}. 
 	During the multi-UE BPL allocation, the network utilizes NZP \mbox{CSI-RS} CM resources for measuring the precoded RSRP/SINR, \mbox{i. e.} using the same precoding as for potential data transmission on the candidate BPL. Dedicated NZP CSI-RS IM resources are used to estimate the interference from other concurrent transmissions via interference-aware CSI-RS scheduling. NZP \mbox{CSI-RSs} can be transmitted in any \ac{OFDM} symbols as configured by the radio resource control, and take fewer radio resources, \mbox{e. g.}~one OFDM symbol, and are transmitted with lower periodicity compared to SSBs~\cite{Zorzi2019_3GPPbeamtutorial}. Step.~3 in Fig.~\ref{fig:SSBCSIRS} shows an example \mbox{MU-MIMO} scenario with three UEs, where one UE receives its NZP \mbox{CSI-RS} for CM while the other two conduct IM at the same time. 
	\textit{IABA} uses the dedicated NZP \mbox{CSI-RS} CM/IM to compute the precoding/combining before each BPL allocation (Step~3 of Alg.~\ref{alg:JointBPL}). At the UE, this corresponds to the beam entry used to steer the array towards the candidate BPL. At the gNB, this is reflected as using a different precoding matrix index for Type II codebooks, which provide multi-beam support suitable for interference cancellation in HBF \mbox{MU-MIMO} configurations with up to four spatial layers, as defined in Sec.~5.2.2.2.1–5 in TS~38.214~\cite{3GPP_NR} and detailed in Sec.~\ref{sec:channel_estimation}--\ref{sec:multiUE_IC}.

\section{Performance Evaluation}
\label{sec:performance_evaluation}

\subsection{Simulation Model}
\label{sec:simulation_model}

\begin{figure}[t!]
	\centering
	\includegraphics[width=0.3\textwidth]{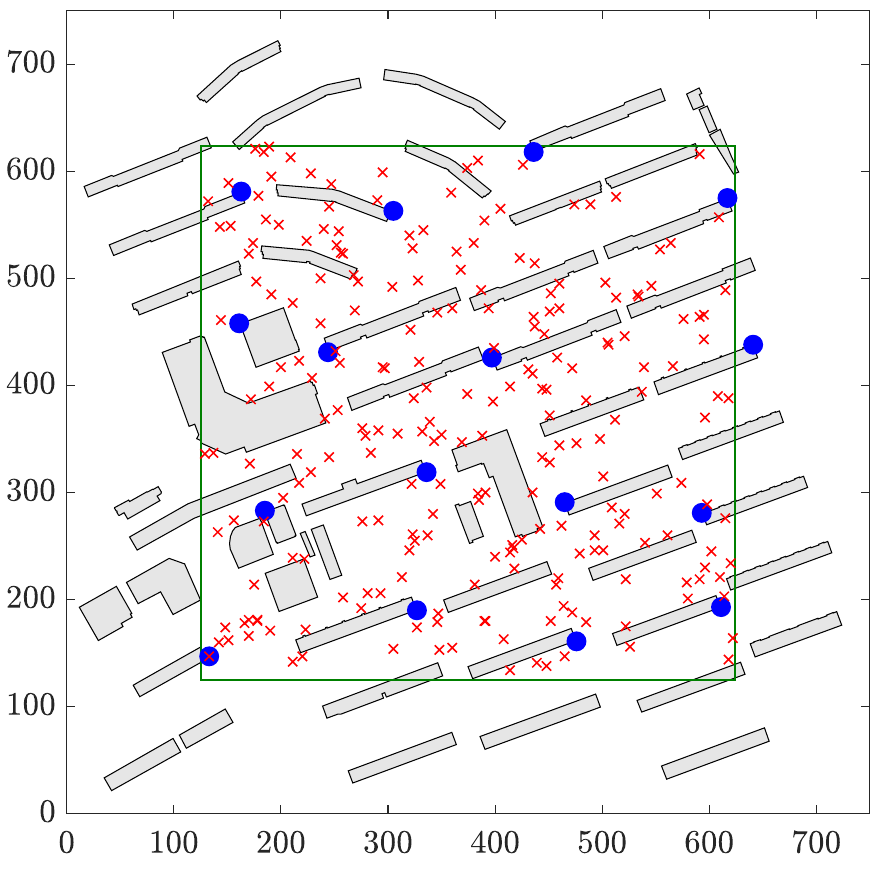}
	\caption{Illustration of the urban site in Seoul, showing the building layout and an example gNB (\textcolor{blue}{$\bullet$}) and UE (\textcolor{red}{$\times$}) deployment over the network study area (green box); $\lambda_{gNB}=64$~$\text{gNBs/km}^2$ and  $\lambda_{UE}=1000$~$\text{UEs/km}^2$.}
   	 \label{fig:0}
\end{figure}

	We consider a realistic urban area in Seoul, with the \mbox{mm-wave} network consisting of gNBs and UEs deployed within a central $500$~$\text{m}$~$\times$~$500~\text{m}$ network area, as shown in Fig.~\ref{fig:0}. The gNB deployment follows a regular distribution on a uniform grid, where the gNB locations are mapped to the closest building corner near the actual grid position. We consider gNB deployment densities of \mbox{$\lambda_{gNB}=\{64, 196\}$~$\text{gNBs/km}^2$}. The sparse deployments are a representative of early mm-wave networks deployed as a hot-spot capacity extension, whereas the dense deployments correspond to future mm-wave networks. The UEs are randomly deployed following a Poisson point process (PPP) with a density of $\mbox{$\lambda_{UE}=1000$~$\text{UEs/km}^2$}$.	
	
	We assume an operating carrier frequency of \mbox{$f_{c}=28$~$\text{GHz}$} and a system bandwidth of $B=400$~MHz, corresponding to \mbox{5G-NR} FR2 network deployments~\cite{etsi_nr}. We consider saturated downlink traffic, and assume a maximum gNB transmit power of \mbox{$P_{max}=30$~$\text{dBm}$} and a noise power of $N=-78$~dBm, consisting of thermal noise over the system bandwidth plus a noise figure of 10~dB.
	We model the gNB antenna as per 3GPP's specifications with a size of up to $N_t=1024$ elements and $N^g_{RF}=4$~RF chains per sector panel~\cite{3GPP_model}, deployed at a height of 6~m. Each gNB is thus equipped with a total of $N^g_{RF}=N_{sec}N^g_{RF,sec}=16$ RF chains. We assume a CBF architecture at the UE and a multi-sector antenna with $N_r=16$ elements per sector panel, deployed at a height of 1.5~m. The UEs are equipped with $N^u_{RF}=1$ RF chain, as per 5G-NR specifications~\cite{3GPP_model}. Throughout, the results represent Monte Carlo simulations with 20 network realizations of random UE deployments. Table~\ref{table:systemParameters} summarizes the studied cellular network parameters.

		\begin{table}[tb]
			\centering
			\small
			\begin{tabular}{l l}
				\toprule
				\textbf{Parameter} & \textbf{Value} \\ 
				\midrule 
				Network area size & $500$~$\text{m}$~$\times$~$500~\text{m}$\\
				gNB distribution & uniform grid \\ 
				gNB density $\lambda_{gNB}$ & $\{64, 196\}$~$\text{gNBs/km}^2$\\ 
				gNB height & $6$~$\text{m}$\\
				UE distribution & PPP\\ 
				UE density $\lambda_{UE}$ & $1000$~$\text{UEs/km}^2$\\ 
				UE height & $1.5$~$\text{m}$\\
				\hline
				\# sector antenna panels $N_{sec}$ & 4 \\
				\# RF chains per gNB sector $N^{g}_{RF,sec}$  & 4 \\
				\# RF chains per gNB  $N^{g}_{RF}$  & 16 \\
				\# RF chains per UE  $N^{u}_{RF}$ & 1 \\
				gNB sector antenna size $N_t$ & \{256, 1024\} elements \\
				UE sector antenna size $N_r$ & 16 elements \\
				\# bits for steering codebook $n_q$ & \{2, 4, 6, 8, 10, $\infty$\} \\
				\# SSBs $N_{SSB}$ & 64 \\
				\# monitored CSI-RS $N_{CSI-RS}$ & \{2, 4, $\infty$\} \\
				\hline		
				Carrier frequency $f_{c}$ & $28$~$\text{GHz}$ \Tstrut \\
				System bandwidth $B$ & $400$~$\text{MHz}$\\
				Max. gNB transmit power $P_{max}$ & $30$~$\text{dBm}$\\
				Noise power $N$ & $-78$~$\text{dBm}$\\	
				Coverage threshold $SINR_{min}$ & $-5$~$\text{dB}$\\
				Max. achievable throughput $R_{max}$ & 2~Gbps \\	
				\bottomrule
		          	\end{tabular}
			\caption{Studied cellular network parameters.} 
			\label{table:systemParameters}
			\vspace{-5pt}
		\end{table}

\subsection{Results}
\label{sec:Results}

	In the following, we present the results of our extensive study of mm-wave HBF MU-MIMO performance and the evaluation of \textit{IABA} under practical network constraints. In Sec.~\ref{sec:baseline}, we study the baseline performance of \textit{IABA} and compare it against default 5G-NR\footnote{In \cite{Ichkov_WCNC23}, we show that IABA outperforms WCS, as a literature benchmark~\cite{Alizadeh_2019} for spatial BPL allocation in CBF mm-wave SDMA networks. We expect the same performance trends to hold, and do not include an explicit comparison against WCS for HBF MU-MIMO in this work. We note that, as far as we are aware of, there are no other \mbox{mm-wave} networking studies on HBF MU-MIMO (\emph{cf.}~Table.~\ref{tab:related_work}) that consider multi-UE interference cancellation and spatial BPL allocation jointly. We therefore select HBF MU-MIMO with default 5G-NR UE association to the strongest gNB and BPL found during initial access (\emph{cf.}~Sec.~\ref{sec:initial_access}) as our main performance evaluation benchmark.} for the ideal case of $n_q=\infty$ and $N_{CSI-RS}=\infty$. 
	In Sec.~\ref{sec:n_csirs} we study the effect of candidate BPL monitoring in \textit{IABA}, and show that \textit{IABA} significantly outperforms default 5G-NR under realistic \mbox{CSI-RS} signaling, \mbox{i. e.} $N_{CSI-RS}\leq4$.
	In Sec.~\ref{sec:ch_quantization} we analyze the effect of quantized CSI feedback, \mbox{i. e.} $n_q<\infty$, on the HBF \mbox{MU-MIMO} performance, showing the benefits of \textit{IABA} and performance gains over the default interference-agnostic 5G-NR. 
	Finally, in Sec.~\ref{sec:4vsInfbitQuan_4vsInfNcsirs} we study the practical performance trade-offs in terms of CSI-RS BPL monitoring $N_{CSI-RS}$ and quantized CSI feedback $n_q$, and show that \textit{IABA} allows HBF MU-MIMO to even outperform the \ac{DBF} MU-MIMO baseline, with further benefits of increased gNB antennas and densification in beyond-5G networks.

\subsubsection{Beyond 5G-NR Baseline Performance}
\label{sec:baseline}
	
\begin{figure*}[tb!]
\centering
\subfloat[SINR]{\includegraphics[width=0.335\textwidth]{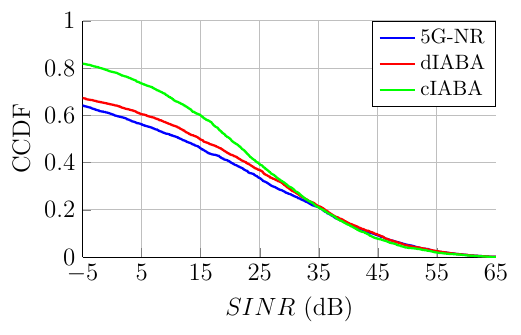}\label{fig:baseline_sinr_hbf}} 
%
\subfloat[INR]{\includegraphics[width=0.335\textwidth]{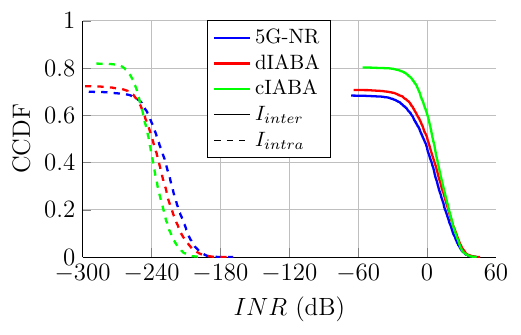}\label{fig:baseline_inr_hbf}}
\subfloat[SNR]{\includegraphics[width=0.335\textwidth]{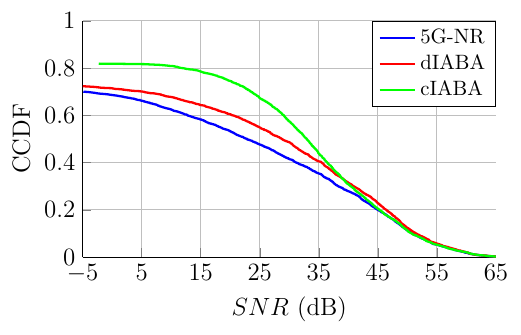}\label{fig:baseline_snr_hbf}} 

\caption{SINR, INR and SNR distribution for the different BPL allocation approaches, assuming $n_q=\infty$, $N_{CSI-RS}=\infty$, \mbox{i. e.} ideal channel estimation and CSI-RS BPL monitoring, and our baseline network of $\lambda_{gNB}=64$~BSs/km$^2$, $\lambda_{UE}=1000$~UEs/km$^2$, $N_t=256/N_r=16$ elements.}
\label{fig:baseline1}
\vspace{-10pt}
\end{figure*}

\begin{figure}[tb!]
\centering
\subfloat[Allocated BPL index]{\includegraphics[width=0.245\textwidth]{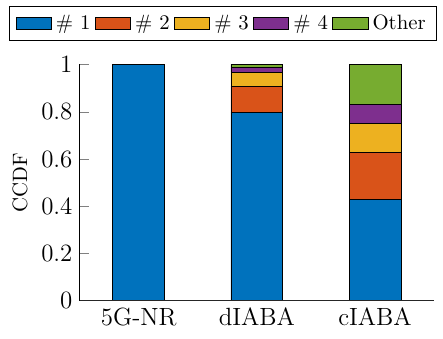}\label{fig:baseline_linkID_hbf}} 
\subfloat[Allocated BPL type]{\includegraphics[width=0.245\textwidth]{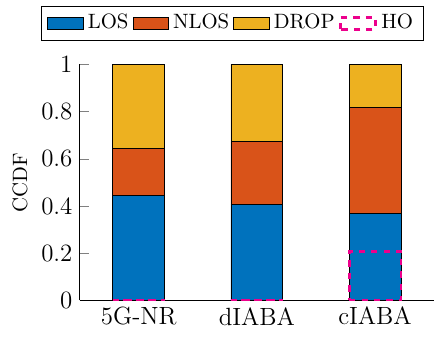}\label{fig:baseline_linkType_hbf}} 
\caption{Index and type of the allocated BPLs, assuming $n_q=\infty$, $N_{CSI-RS}=\infty$, $\lambda_{gNB}=64$~BSs/km$^2$, $\lambda_{UE}=1000$~UEs/km$^2$, and $N_t=256/N_r=16$ elements.}
\label{fig:baseline2}
\vspace{-10pt}
\end{figure}

	Let us being by studying the performance of \textit{IABA}, and compare it against the interference-agnostic, default 5G-NR BPL allocation (referred to as \textit{5G-NR}) for our baseline network deployment of \mbox{$\lambda_{gNB}=64$~BSs/km$^2$}, \mbox{$\lambda_{UE}=1000$~UEs/km$^2$}, and \mbox{$N_t=256/N_r=16$} elements.  We additionally assume ideal channel estimation and \mbox{CSI-RS} BPL monitoring, i. e. $n_q=\infty$ and $N_{CSI-RS}=\infty$.
	Fig.~\ref{fig:baseline1} presents the resulting \ac{SINR}, \ac{INR} and \ac{SNR} distributions and Fig.~\ref{fig:baseline2} illustrates the network BPL allocation, i.e. index/type of allocated BPLs, for the different BPL allocation schemes.
	
	Fig.~\ref{fig:baseline_sinr_hbf} shows that the proposed \textit{IABA} significantly improves both the coverage and achieved SINR over the \mbox{\textit{5G-NR}} baseline. The centralized approach \textit{cIABA} improves the coverage by 20 percentage points (pp), and the median SINR by 10~dB over \textit{5G-NR}. The distributed approach \textit{dIABA} also outperforms \mbox{\textit{5G-NR}} by 4~pp in terms of coverage and over 4~dB in the median SINR.  
	Interestingly, while Fig.~\ref{fig:baseline_inr_hbf} confirms that zero-forcing in HBF MU-MIMO fully cancels the \textit{intra-cell} interference, there is a significant portion of allocated UEs that are still affected by strong residual \mbox{\textit{inter-cell}} interference, i. e. 60\% of the UEs for \textit{cIABA} and over 45\% for \textit{dIABA}/\mbox{\textit{5G-NR}} experience inter-cell interference above the noise floor ($INR > 0$). We emphasize that this result is in contrast to assumptions made in prior \mbox{mm-wave} studies, which typically assume limited to no-interference due to simplified antenna/channel models, e.g.~\cite{Kwon2019_SDMA, Alizadeh_2019, Alizadeh_COMCAS2019, NIF2022, versatileRRM2023}. 		
	
	Let us further analyze the nature of the residual spatial interference in HBF MU-MIMO mm-wave networks to gain insight into how \textit{IABA} achieves its performance gains over the \textit{5G-NR} baseline in Fig.~\ref{fig:baseline_sinr_hbf}. Zero-forcing cancels intra-cell interference by projecting the UE's channel vector on a precoding vector that is orthogonal to other UEs' channel vectors~(\ref{eq:zero_forcing}). The penalty for this depends on the mutual channel correlation, and is reflected as a loss in the \ac{RSS} if the channels are highly correlated, known as the \mbox{zero-forcing} penalty~\cite{Kularni2016-MIMOComparison}. The \mbox{mm-wave} channel is sparse, and the effective UE channel depends on the choice of the beam codebook entries which maximize the RSS of the candidate BPLs found during initial access. In practice, the chosen beams amplify certain channel paths and attenuate others, so that the allocated BPLs in the \textit{5G-NR} baseline typically share common strong paths, \mbox{e. g.} for nearby \mbox{co-located} UEs, which results in high channel correlation. In such cases, the zero-forcing penalty incurs significant losses in order to fully cancel intra-cell interference, due to its inability to compensate for the low angular separation of the allocated BPLs. Namely, always selecting the best RSS BPLs at initial access for serving all UEs as in \mbox{\textit{5G-NR}} results in a high zero-forcing penalty and thus poor SNR after zero-forcing precoding, as shown in Fig.~\ref{fig:baseline_snr_hbf}. By contrast, the proposed \textit{IABA} is able to significantly improve the SNR in Fig.~\ref{fig:baseline_snr_hbf} by effectively increasing the angular separation of the allocated BPLs after zero-forcing via selection of the $n^\text{th}$-strongest BPLs at initial access, as shown in Fig.~\ref{fig:baseline_linkID_hbf}.		
	By allocating secondary BPLs to increase the angular BPL separation, \mbox{i.~e.~$n > 1$}, \textit{IABA} reduces the zero-forcing penalty and mitigates strong interference that would otherwise result by only allocating the best RSS BPL, \mbox{i.~e.~$n = 1$}. For example, 20\% of the UEs for \textit{dIABA} and 57\% for \mbox{\textit{cIABA}} are allocated BPLs other than the best RSS BPL, as shown in Fig.~\ref{fig:baseline_linkID_hbf}. Moreover, Fig.~\ref{fig:baseline_linkType_hbf} shows that 45\% of the UEs are allocated NLOS BPLs for \mbox{\textit{cIABA}}, compared to 27\% for \textit{dIABA} and only 19\% for \mbox{\textit{5G-NR}}.
\mbox{\textit{cIABA}} further improves the overall network performance by allowing handovers for 21\% of the UEs in Fig.~\ref{fig:baseline_linkType_hbf}, thus having the benefit of load-balancing and outperforming the other BPL allocation approaches. 

	In summary, the results in Figs.~\ref{fig:baseline1}-\ref{fig:baseline2} demonstrate the significant benefit of interference-aware BPL allocation to manage the residual spatial interference in HBF MU-MIMO \mbox{mm-wave} networks that cannot be mitigated by zero-forcing alone even under ideal channel estimation ($n_q=\infty$).

\subsubsection{CSI-RS BPL Monitoring for IABA}
\label{sec:n_csirs}

\begin{figure*}[t!]
\centering
\subfloat[SINR]{\includegraphics[width=0.335\textwidth]{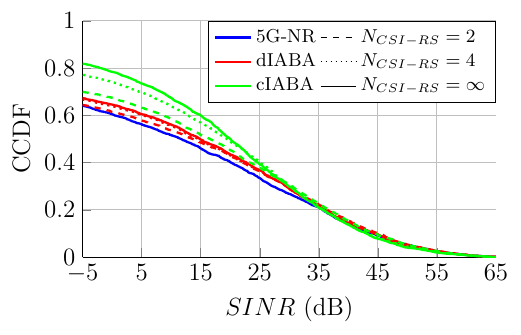}\label{fig:sinr_hbf_Ncsirs}} 
\subfloat[Allocated BPL index]{\includegraphics[width=0.335\textwidth]{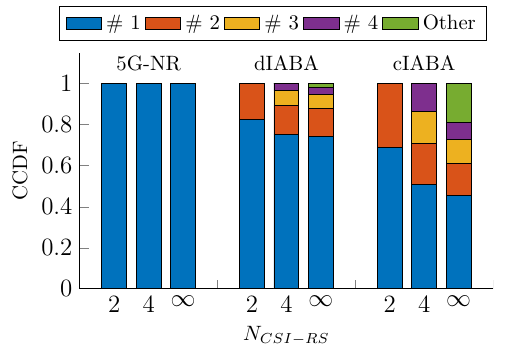}\label{fig:linkID_hbf}} 
\subfloat[Allocated BPL type]{\includegraphics[width=0.335\textwidth]{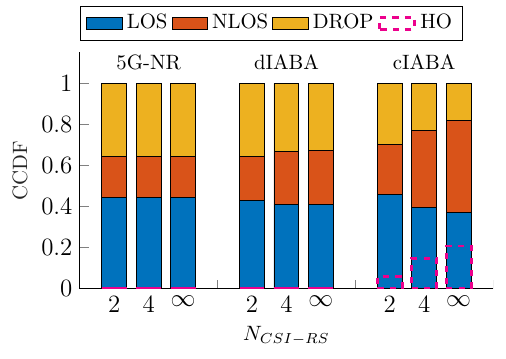}\label{fig:linkType_hbf}} 

\caption{SINR distribution and allocated BPL index/type for varying $N_{CSI-RS}$, assuming ideal channel estimation $n_q=\infty$, $\lambda_{gNB}=64$~BSs/km$^2$, $\lambda_{UE}=1000$~UEs/km$^2$, and $N_t=256/N_r=16$ elements.}
\vspace{-18pt}
\label{fig:Ncsirs}
\end{figure*}

	Let us next focus on how \textit{IABA} achieves network-wide performance gains in practice for our baseline network deployment. Fig.~\ref{fig:Ncsirs} shows the SINR distribution and the allocated BPL index/type for varying \mbox{$N_{CSI-RS}=\{2, 4, \infty\}$} under ideal channel estimation \mbox{($n_q=\infty$)}. 
	Fig.~\ref{fig:sinr_hbf_Ncsirs} shows that constraining $N_{CSI-RS}$ has only a limited impact on the achieved SINR for \textit{dIABA}. This is because the ideal allocation ($N_{CSI-RS}=\infty$) of the per-UE BPLs for \textit{dIABA} picks at most the second best BPL for 88\% of UEs, and 98\% up to the fourth best BPL, as shown in Fig.~\ref{fig:linkID_hbf}. Limiting $N_{CSI-RS}$ to four thus results in no significant impact of the achievable network performance for \textit{dIABA} in Fig.~\ref{fig:sinr_hbf_Ncsirs}. 
	
	In contrast, Fig.~\ref{fig:linkID_hbf} shows that \textit{cIABA} allocates the fifth or higher BPL index for 19\% of the UEs for $N_{CSI-RS}=\infty$. Importantly, only 45\% of the UEs are allocated the best RSS BPL for $N_{CSI-RS}=\infty$, compared to 50\% for $N_{CSI-RS}=4$ and 68\% for $N_{CSI-RS}=2$. Thus, limiting $N_{CSI-RS}$ for \textit{cIABA} has a more significant impact in terms of SINR degradation in Fig.~\ref{fig:sinr_hbf_Ncsirs} than for \textit{dIABA}. Reducing $N_{CSI-RS}$ also limits the use of secondary NLOS BPLs for interference mitigation in \textit{cIABA}, and increases the primary use of LOS BPLs, \mbox{e. g.} from 37\% for $N_{CSI-RS}=\infty$ up to 46\% for $N_{CSI-RS}=2$ in Fig.~\ref{fig:linkType_hbf}. In addition, the load-balancing rate via handovers drops from 21\% down to 5\% for $N_{CSI-RS}=2$ in Fig.~\ref{fig:linkType_hbf}. Consequently, Fig.~\ref{fig:sinr_hbf_Ncsirs} shows that \textit{cIABA} with $N_{CSI-RS}=2$ achieves 15~pp lower coverage than \textit{cIABA} with $N_{CSI-RS}=\infty$, with its overall SINR performance being comparable to \textit{dIABA} with $N_{CSI-RS}=\infty$.
	
	Our results in Fig.~\ref{fig:Ncsirs} thus highlight that CSI-RS BPL monitoring improves both coverage and achievable SINR, and is particularly important for \textit{cIABA}. Limiting the number of monitored candidate BPLs inherently limits the load-balancing opportunities for \textit{cIABA}, considering that the best candidate BPLs found during initial access are provided by the initial serving gNB. Nonetheless, \textit{cIABA} monitoring up to the $N_{CSI-RS}=4$ best candidate BPLs -- in line with 3GPP's specifications for FR2~\cite{Zorzi2019_3GPPbeamtutorial} -- is able to reap the benefits of IABA's CSI-RS BPL monitoring and significantly outperform \mbox{\textit{5G-NR}} in Fig.~\ref{fig:sinr_hbf_Ncsirs}, by allocating secondary BPLs to over 50\% of the UEs (\emph{cf.}~Fig.~\ref{fig:linkID_hbf}). This result demonstrates the practical feasibility of \textit{IABA} for network-wide, interference-aware BPL allocation in HBF MU-MIMO networks.

\subsubsection{CSI Reporting \& Channel Feedback Quantization}
\label{sec:ch_quantization}

\begin{figure*}[tb!]
\centering
\subfloat[SINR]{\includegraphics[width=0.335\textwidth]{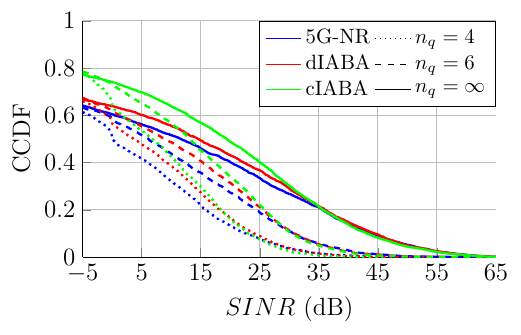}\label{fig:sinr_bitQuan_hbf}}
\subfloat[$I_{intra}$NR]{\includegraphics[width=0.335\textwidth]{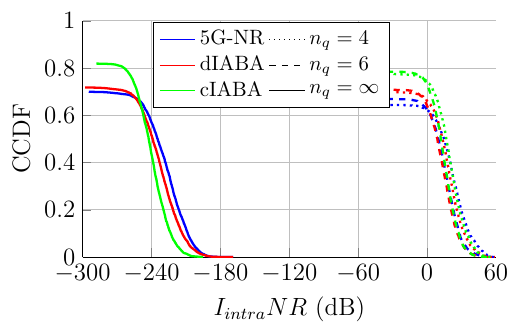}\label{fig:intraCell_bitQuan_hbf}} 
\subfloat[$I_{inter}$NR]{\includegraphics[width=0.335\textwidth]{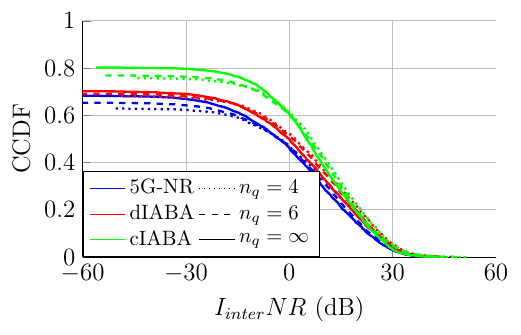}\label{fig:interCell_bitQuan_hbf}} \\

\caption{SINR and INR distribution using quantized channel feedback from varying codebook sizes $n_q=\{4, 6, \infty\}$,   assuming $N_{CSI-RS}=4$, $\lambda_{gNB}=64$~BSs/km$^2$, $\lambda_{UE}=1000$~UEs/km$^2$, and $N_t=256/N_r=16$ elements.}
\label{fig:baseline_bitQuan}
\vspace{-10pt}
\end{figure*}
\begin{figure}[tb!]
\centering
\subfloat[Angular misalignment $\Delta \phi$]{\includegraphics[width=0.245\textwidth]{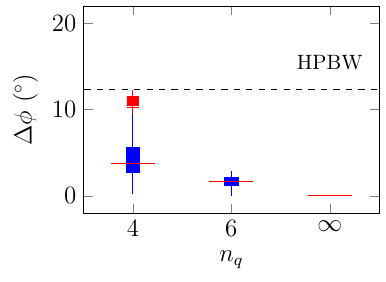}\label{fig:misalignmentDegrees_bitQuan_hbf}} 
\subfloat[Gain loss $\Delta G$]{\includegraphics[width=0.245\textwidth]{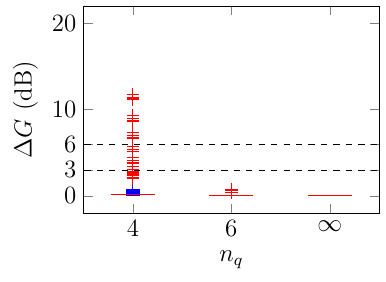}\label{fig:misalignmentLoss_bitQuan_hbf}} 

\caption{Distribution of angular misalignment between the steering orientation of the codebook main lobe and the ideal BPL orientation $\Delta \phi$, and the incurred gain loss due to misalignment $\Delta G$, assuming codebook sizes of $n_q=\{4, 6, \infty\}$ for our baseline network deployment.}
\label{fig:misalignment_bitQuan}
\vspace{-10pt}
\end{figure}

	Next, we analyze the performance of \textit{IABA} assuming quantized channel feedback for our baseline HBF MU-MIMO network deployment. We set $N_{CSI-RS}=4$ as per Sec.~\ref{sec:n_csirs}, and vary the size of the codebooks used for channel estimation, $n_q=\{4, 6, \infty\}$, as discussed in detail in Sec.~\ref{sec:channel_estimation}. 	

	Fig.~\ref{fig:baseline_bitQuan} shows the resulting SINR and INR distributions using quantized channel feedback from varying codebook sizes. Fig.~\ref{fig:sinr_bitQuan_hbf} shows that finite codebooks can achieve good coverage, but lead to significant performance degradation, \mbox{i. e.} over 10~dB loss in the median SINR for the smallest $n_q=4$ codebook compared to the ideal $n_q=\infty$ codebook. Fig.~\ref{fig:sinr_bitQuan_hbf} shows that a codebook of size $n_q=6$ bits still suffers from significant SINR loss compared to the ideal $n_q=\infty$ codebook, due to the inability to perform ideal zero-forcing given the quantized channel estimates.
We emphasize that this is despite the fact that a codebook of size $n_q=\{4, 6\}$ bits does not incur significant angular misalignment between the steering orientation of the main lobe for the used codebook entry and the ideal BPL orientation corresponding to the true channel propagation path, i. e. the misalignment is within the \ac{HPBW} in Fig.~\ref{fig:misalignmentDegrees_bitQuan_hbf}. In particular, Fig.~\ref{fig:misalignmentLoss_bitQuan_hbf} shows that a codebook of size $n_q=6$~bits already performs very close to the ideal in terms of incurred gain loss due to misalignment. Instead the detrimental effect of the quantized codebook comes from poor intra-cell interference suppression due to inaccurate channel estimates. 

	Fig.~\ref{fig:intraCell_bitQuan_hbf} confirms that, unlike the ideal case of $n_q=\infty$, the HBF MU-MIMO network is unable to fully cancel the intra-cell interference using a finite resolution codebook, and there is strong residual intra-cell interference affecting over 60\% of UEs for $n_q=\{4, 6\}$, regardless of the BPL allocation approach. On the other hand, Fig.~\ref{fig:interCell_bitQuan_hbf} shows that the extent of inter-cell interference is not strongly affected as long as the codebook steering resolution covers the \ac{HPBW}, as is the case for both codebooks of $n_q=\{4, 6\}$. Fig.~\ref{fig:intraCell_bitQuan_hbf} thus highlights that practical limitations on the channel estimation codebook size and resulting accuracy of channel estimates strongly limit the extent to which spatial interference in HBF \mbox{MU-MIMO} networks can be mitigated by interference cancellation techniques such as zero-forcing. 
	This emphasizes the even greater importance in practical networks of additional interference mitigation via spatial separation of BPL allocation for beyond-5G deployments. Fig.~\ref{fig:baseline_bitQuan} shows these benefits of interference-aware BPL allocation for practical deployments: \textit{IABA} significantly improves coverage and SINR over \mbox{\textit{5G-NR}} under quantized CSI feedback $n_q=\{4, 6\}$ and realistic \mbox{CSI-RS} BPL monitoring $N_{CSI-RS}=4$.

\subsubsection{Towards Practical Interference-aware BPL Allocation for Beyond-5G}
\label{sec:4vsInfbitQuan_4vsInfNcsirs}
	
	Finally, let us analyze the practical performance trade-offs in terms of channel estimation, CSI availability, and BPL monitoring, towards the feasibility of interference-aware BPL allocation as the proposed \textit{IABA} for \mbox{beyond-5G} \mbox{mm-wave} networks. We consider codebook sizes of \mbox{$n_q=\{4, 6\}$}~bits as per Sec.~\ref{sec:ch_quantization}, and set the number of monitored \mbox{CSI-RS} BPLs \mbox{$N_{CSI-RS}=4$} as per Sec.~\ref{sec:n_csirs}. We then compare HBF \mbox{MU-MIMO} with different BPL allocation approaches against two reference mm-wave candidate architectures, both with default \textit{5G-NR} BPL allocation: 
	\begin{enumerate}[label=\emph{(\roman*)}]
	\item DBF MU-MIMO: UEs are served via SDMA using DBF and zero-forcing for multi-UE interference cancellation, following the same system model considerations as for HBF MU-MIMO in~Sec.~\ref{sec:antenna_BPL_model}--\ref{sec:channel_estimation},
	\item CBF SU-MIMO: UEs are served via TDMA using CBF, corresponding to current 5G-NR FR2 networks.
	\end{enumerate}

\begin{figure*}[t!]
\centering
\hspace{20pt}
\includegraphics[width=0.88\textwidth]{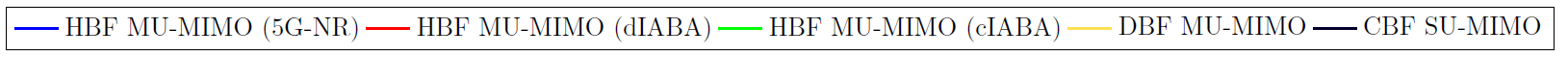} 
\vspace{-12pt}
	
\centering
\subfloat[SINR ($n_q=\infty, N_{CSI-RS}=\infty$)]{\includegraphics[width=0.335\textwidth]{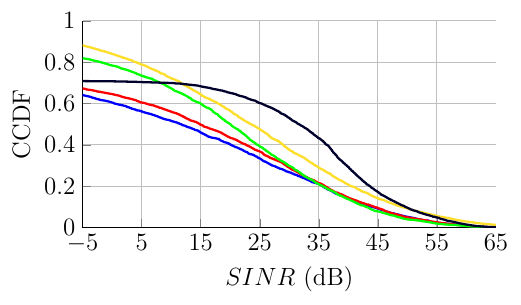}\label{fig:sinr_quan_ideal}} 
\subfloat[SINR ($n_q=6, N_{CSI-RS}=4$)]{\includegraphics[width=0.335\textwidth]{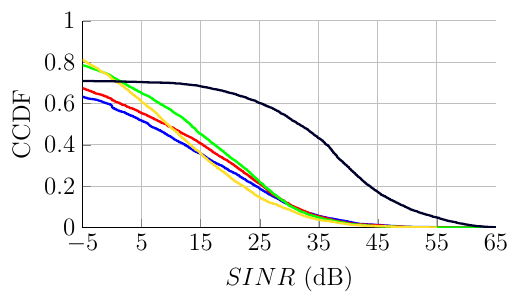}\label{fig:sinr_quan6}}
\subfloat[Throughput ($n_q=6, N_{CSI-RS}=4$)]{\includegraphics[width=0.335\textwidth]{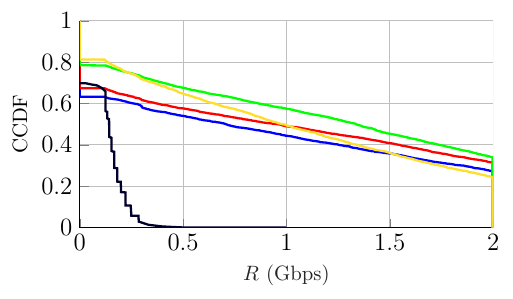}\label{fig:rate_quan6}} 


\caption{SINR and throughput distribution for different beyond-5G network configurations and practical channel estimation and CSI-RS reporting ($\lambda_{gNB}=64$~BSs/km$^2$, $\lambda_{UE}=1000$~UEs/km$^2$, and $N_t=256/N_r=16$ elements).}
\label{fig:quan_peralloc}
\vspace{-10pt}
\end{figure*}

	We start with the reference network performance under ideal conditions ($n_q=\infty$, $N_{CSI-RS}=\infty$) for our baseline network with $\lambda_{BS}=64$~BSs/km$^2$, $\lambda_{UE}=1000$~UEs/km$^2$, and $N_t=256/N_r=16$ elements. Fig.~\ref{fig:sinr_quan_ideal} presents the SINR distribution, confirming that DBF MU-MIMO performs the best, whereas CBF SU-MIMO is able to provide high SINR at the cost of reduced coverage. Fig.~\ref{fig:sinr_quan_ideal} also shows that HBF MU-MIMO with default \textit{5G-NR} BPL allocation performs the worst, with the proposed \textit{IABA} enabling HBF MU-MIMO to close the gap to the DBF MU-MIMO baseline. In particular, Fig.~\ref{fig:sinr_quan_ideal} shows that HBF MU-MIMO with \textit{cIABA} performs within 3.5~dB of the DBF MU-MIMO baseline for the median SINR, and significantly outperforms the default \textit{5G-NR} HBF \mbox{MU-MIMO}, e. g. 8.5~dB gain for the median SINR case. 
	
	Let us now move to the more practical network deployments, with Fig.~\ref{fig:sinr_quan6} showing the SINR distribution for the case of $n_q=6$~bits and $N_{CSI-RS}=4$. Comparing Figs.~\ref{fig:sinr_quan_ideal} and \ref{fig:sinr_quan6} confirms that both HBF and DBF \mbox{MU-MIMO} suffer from significant performance degradation due to imperfect interference cancellation using quantized channel estimates, whereas CBF \mbox{SU-MIMO} performance is robust since the misalignment loss of the steering codebook is negligible (\emph{cf.}~Fig.~\ref{fig:misalignment_bitQuan}). 
Importantly, Fig.~\ref{fig:sinr_quan6} shows that HBF MU-MIMO benefits strongly from the proposed \textit{IABA}, and is able to outperform DBF over the complete SINR range. 
This is also because DBF is particularly affected under quantized CSI feedback, since the accuracy of the estimated channel matrix is significantly degraded due to the higher dimensionality requirements, degrading the multi-UE interference cancellation performance. Namely, HBF \mbox{MU-MIMO} is affected by a quantized codebook to a lesser extent than DBF, due its lower requirements on the effective UE channel estimates. 
To better understand the network performance, Fig.~\ref{fig:rate_quan6} maps using (\ref{eq:rate}) the SINR in Fig.~\ref{fig:sinr_quan6} to the achievable per-UE throughput for the different configurations. Fig.~\ref{fig:rate_quan6} shows that HBF MU-MIMO with \textit{cIABA} outperforms all other configurations, and achieves a median throughput of $1.35$~Gbps, which is 35\% higher than that of the DBF \mbox{MU-MIMO} baseline. Fig.~\ref{fig:rate_quan6} also shows that current \mbox{5G-NR} deployments relying on CBF \mbox{SU-MIMO} are only able to achieve limited per-UE throughput when a large number of UEs are served on each gNB, due to the time resource sharing of TDMA. Furthermore, Fig.~\ref{fig:rate_quan6} shows that HBF \mbox{MU-MIMO} utilizing \textit{dIABA} achieves better performance for over half of the UEs compared to the DBF \mbox{MU-MIMO} baseline, though at a cost of somewhat reduced coverage; this makes the less computationally demanding \textit{dIABA} particularly suitable for dense mm-wave SDMA deployments.

\begin{figure}[t!]
\centering
\hspace{15pt}
\includegraphics[width=0.215\textwidth]{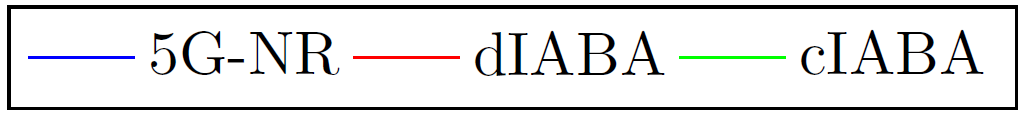} 
\vspace{-12pt}

\centering
\subfloat[Varying $N_t$ ($\lambda_{gNB}=64$~gNBs/km$^2$).]{\includegraphics[width=0.345\textwidth]{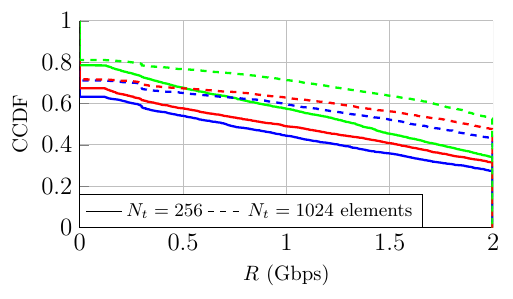}\label{fig:rate_quan_antenna}} \\
\vspace{-10pt}
\subfloat[Varying $\lambda_{gNB}$ ($N_t=256$ elements).]{\includegraphics[width=0.345\textwidth]{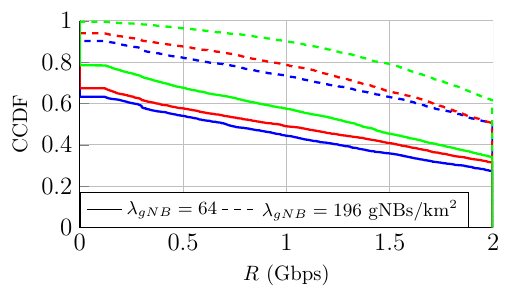}\label{fig:rate_quan_density}} 

\caption{Throughput distribution for the different BPL allocation approaches, assuming HBF MU-MIMO with varying gNB antenna $N_t$ and deployment density $\lambda_{gNB}$ ($\lambda_{UE}=1000$~BS/km$^2$, $N_r=16$ elements).}
\label{fig:rate_quan_scalability}
\vspace{-10pt}
\end{figure}

	Our results in Fig.~\ref{fig:quan_peralloc} thus highlight that the move towards fully-scaled SDMA necessitates interference-aware BPL allocation in order to unlock the full potential of mm-wave for beyond-5G networks. With the envisioned antenna size increase and network densification, we lastly look at the scalability of \textit{IABA} in practical mm-wave HBF \mbox{MU-MIMO} networks. In Fig.~\ref{fig:rate_quan_scalability} we consider the case when UEs are equipped \mbox{$N_r=16$} elements per sector antenna, and present the throughput distribution for a larger gNB sector antenna of \mbox{$N_t=1024$} elements and a higher gNB density of $\lambda_{gNB}=196$~gNBs/km$^2$, alongside our baseline network deployment of \mbox{$N_t=256/\lambda_{gNB}=64$} as shown in Fig.~\ref{fig:rate_quan6}.
		
	Fig.~\ref{fig:rate_quan_antenna} highlights a strong benefit from increased gNB sector antenna for all BPL allocation approaches. The proposed \textit{IABA} yields significant performance gains when moving from a 256- to 1024-element sector antenna: a two-fold increase in the median throughput for \textit{dIABA} from 950~Mbps to 1.88~Gbps, and a similar increase for \textit{cIABA} from 1.3~Gbps to 2~Gbps, as show in Fig.~\ref{fig:rate_quan_antenna}. The increased gNB antenna directionality makes more candidate BPLs feasible for allocation, thus increasing the gNB's ability to spatially distinguish nearby UEs and thus increase the co-scheduled UE BPL separation. For example, more than half of the UEs get the maximum achievable throughput of 2~Gbps with HBF MU-MIMO using \textit{cIABA}, amounting to a 10 pp increase over the interference-agnostic \textit{5G-NR} baseline. Fig.~\ref{fig:rate_quan_antenna} thus shows that large antenna arrays in fact benefit \textit{more} from interference-aware BPL allocation, further highlighting the limiting factor of spatial interference in realistic \mbox{mm-wave} SDMA deployments. 
	
	Fig.~\ref{fig:rate_quan_density} presents the effect of network densification as another crucial aspect towards improving the wide-area \mbox{mm-wave} coverage and achievable performance, assuming \mbox{$N_t=256$} elements per gNB sector. The denser deployment increases the ability of $\textit{dIABA}$ and \textit{cIABA} to mitigate interference, and in particular, improves the load-balancing effect of \textit{cIABA}. Fig.~\ref{fig:rate_quan_density} shows that the three-fold increase in gNB deployment density results in a significant increase in the per-UE achievable throughput -- over 60\% of UEs get the maximum achievable throughput using \textit{cIABA}, compared to the 35\% for the sparse deployment. Similar trends are observed for both \textit{dIABA} and \textit{5G-NR}. Importantly, Fig.~\ref{fig:rate_quan_density} also shows that \textit{cIABA} achieves full network-wide coverage for the dense deployment, and a 10 pp improvement in coverage compared to the \textit{5G-NR} baseline. 
This highlights the primary importance of interference-aware BPL allocation for mm-wave SDMA in dense network deployments, thus validating the design strategy of the proposed \textit{IABA} for practical beyond-5G HBF \mbox{MU-MIMO} networks.

\section{Conclusions}
\label{sec:Conclusions}
	We proposed \textit{IABA}, a 5G-NR standard-compliant BPL allocation scheme, for mitigating residual spatial interference in practical mm-wave HBF MU-MIMO networks. 
	\textit{IABA} solves the network sum throughput optimization via either a \textit{distributed} or a \textit{centralized} BPL allocation using dedicated \ac{CSI-RS} resources for candidate BPL monitoring. 
	We presented a comprehensive study of practical multi-cell \mbox{mm-wave} HBF MU-MIMO networks, and showed that spatial interference becomes the main limiting factor of HBF \mbox{MU-MIMO} performance, which is in contrast to prior mm-wave networking studies assuming simplified channel and antenna models and perfect CSI at the gNB. 
	Our results showed that \textit{IABA} offers significant performance gains over default \mbox{5G-NR}, by facilitating allocation of secondary BPLs other than the strongest BPL found during initial access. Importantly, we demonstrated that \textit{IABA} allows HBF \mbox{MU-MIMO} to outperform the interference-agnostic \ac{DBF} MU-MIMO baseline, with further benefits of increased gNB antennas and densification for beyond-5G mm-wave networks.
	Our ongoing work is focused on low-complexity BPL allocation schemes using site-specific codebooks for adaptive CSI acquisition and further spatial multiplexing gains in diverse network configuration deployments. 
	
\ifCLASSOPTIONcaptionsoff
  \newpage
\fi


\bibliographystyle{IEEEtran}
\bibliography{bibliography}

\end{document}